\documentclass[doublespacing]{bmcart}

\usepackage[utf8]{inputenc} 

\usepackage{amsmath}
\usepackage{amsfonts}
\usepackage{amssymb}
\usepackage{graphicx}
\usepackage[skip=2pt,font=scriptsize]{caption}
\usepackage{subcaption}

\usepackage{bm} 
\usepackage[pagewise]{lineno}
\usepackage{color}
\usepackage{changes}
\usepackage[normalem]{ulem}
\usepackage{hyperref}

\usepackage{comment}

\usepackage{algorithm}
\usepackage[noend]{algpseudocode}
\usepackage{booktabs} 
\usepackage{multirow}
\usepackage[font={bf},textfont=md]{caption}

\usepackage{etoolbox}
\makeatletter
\patchcmd\set@numberlines@box{\rlap}{\@gobble}{}{}
\makeatother

\usepackage{xcolor}

\startlocaldefs
\endlocaldefs

\begin{document}

\begin{frontmatter}

\begin{fmbox}
\dochead{Research}


\title{Understanding vehicular routing behavior with location-based service data}


\author[
   addressref={aff1,aff2,aff3,aff4},                   
   corref={aff1},                       
   email={yanyanxu@berkeley.edu}   
]{\inits{YX}\fnm{Yanyan} \snm{Xu}}
\author[
   addressref={aff2,aff5,aff6},
   email={rr.di-clemente@exeter.ac.uk}
]{\inits{RDC}\fnm{Riccardo Di } \snm{Clemente}}
\author[
   addressref={aff2,aff3,aff4,aff7},
   email={martag@berkeley.edu}
]{\inits{MCG}\fnm{Marta C.} \snm{Gonz\'{a}lez}}


\address[id=aff1]{
  \orgname{MoE Key Laboratory of Artificial Intelligence, AI Institute, Shanghai Jiao Tong University}, 
  \postcode{200240},                                
  \city{Shanghai},                              
  \cny{China}                                    
}

\address[id=aff2]{
  \orgname{Department of Civil \& Environmental Engineering, MIT}, 
  \postcode{02139},                                
  \city{Cambridge, MA},                              
  \cny{USA}                                    
}
\address[id=aff3]{%
  \orgname{Department of City and Regional Planning, University of California, Berkeley},
  \postcode{94720}
  \city{Berkeley, CA},
  \cny{USA}
}
\address[id=aff4]{%
  \orgname{Energy Analysis \& Environmental Impacts Division, Lawrence Berkeley National Laboratory},
  \postcode{94720}
  \city{Berkeley, CA},
  \cny{USA}
}
\address[id=aff5]{%
  \orgname{Centre for Advanced Spatial Analysis, University College London},
  \city{London WC1E 6BT},
  \cny{UK}
}
\address[id=aff6]{%
  \orgname{Department of Computer Science, University of Exeter},
  \city{Exeter, EX4 4NR},
  \cny{UK}
}
\address[id=aff7]{%
  \orgname{Department of Civil and Environmental Engineering, University of California, Berkeley},
  \postcode{94720},
  \city{Berkeley, CA},
  \cny{USA}
}



\end{fmbox}


\begin{abstractbox}

\begin{abstract} 
Properly extracting patterns of individual mobility with high resolution data sources such as the one extracted from smartphone applications offers important opportunities. Potential opportunities not offered by call detailed records (CDRs), which offer resolutions triangulated from antennas, are route choices, travel modes detection and close encounters. Nowadays, there is not a standard and large scale data set collected over long periods that allows us to characterize these. In this work we thoroughly examine the use of data from smartphone applications, also referred to as location-based services (LBS) data, to extract and understand the vehicular route choice behavior. Taking the Dallas-Fort Worth metroplex as an example, we first extract the vehicular trips with simple rules and reconstruct the origin-destination matrix by coupling the extracted vehicular trips of the active LBS users and the United States census data. We then present a method to derive the commonly used routes by individuals from the LBS traces with varying sample rate intervals. We further inspect the relation between the number of routes and the trip characteristics, including the departure time, trip length and travel time. Specifically, we consider the travel time index and buffer index for the LBS users taking different number of routes. Empirical results demonstrate that during the peak hours, travelers tend to reduce the impact of traffic congestion by taking alternative routes. 
Overall, the proposed data analysis framework is cost-effective to treat sparse data generated from the use of smartphones to inform routing behavior. The potential in practice is to inform demand management strategies, by targeting individual users while generating large scale estimates of congestion mitigation.
\end{abstract}



\linenumbers


\end{abstractbox}
%

\end{frontmatter}



\section{Introduction}
\label{sec:intro}
With the growing population in cities and the restructuring of urban economies and societies, a fundamental task of transportation planners and engineers is to effectively move people and goods~\cite{xu2020deconstructing}. However, due to the daily increasing contradiction between the travel demand of citizens and the limit of road resources, the worsening traffic congestion not only causes tremendous economic loss and environmental problems but also has profound impacts on the public health~\cite{weisbrod2003measuring, jiang2017transport, xu2019unraveling}. Today, a number of strategies have been proposed by policymakers and researchers over the world to relieve the traffic congestion, including advanced traffic signal control~\cite{hu1997day}, strengthening the public transportation~\cite{kelly2016transport}, reducing the travel demand of private vehicles~\cite{xu2017collective}, route recommendations~\cite{ccolak2016understanding}, congestion pricing~\cite{prud2005london}, or even looking into the future on autonomous vehicles~\cite{wu2018stabilizing}.

Among the varying travel management strategies, understanding human mobility is always a fundamental task, supporting advanced decision systems. Thanks to the development of modern information and communication technologies (ICT) and the high penetration rate of mobile phone devices, researches can leverage on a large amount of digital traces with time stamps and geographical locations to understand and reproduce human mobility~\cite{gonzalez2008understanding, pappalardo2015returners, jiang2016timegeo}. Despite daily destinations of human mobility can be modeled at half square kilometer scale, the routing behavior of travelers in their road networks is still not well modeled from ICT data sources~\cite{ben1999discrete}. A general solution to the lack of complete information is leveraging traffic assignment models, such as the user equilibrium assignment, dynamic traffic assignment, and the multi-agent approach, to assign each traveler to specific routes~\cite{prato2009route}. These models assume that travelers choose their routes with the intention of minimizing travel costs, such as travel distance or time. To this end, individuals are assumed to have perfect or partial information about the alternatives available to them. However, the choice of route is simultaneously affected by multiple factors, and the route choice behavior of people follows the bounded rationality principle~\cite{di2016boundedly}. This means that travelers can neither find the optimal routes because of the lack of accurate information about the traffic conditions, nor willing to spend much effort to obtain the optimized decision from complicated situations~\cite{zhu2015people}. With vehicular trajectories collected periodically from hundreds of travelers during several months, starting when the driver turns the engine on, until it is turned off, Lima et al. found that the individual routing behavior is independent of the urban layout. People always have a dominant route and the alternative routes are bounded within an elliptic shape of high eccentricity~\cite{lima2016understanding}. In this work, we attempt similar analysis with the additional challenge imposed by data with less temporal accuracy and much lower frequency of collection, but much more pervasive than the vehicular trajectory data.

The prevailing big data resources utilized to model travel behavior include mobile phone data, check-in data at places of interest, and data collected by transportation agencies, like the floating car data. Among them, the mobile phone data, a.k.a. call detail records (CDRs), are passively collected and have the largest coverage. CDRs have been used to understand the mobility behavior~\cite{di2018sequences}, and reproduce the aggregated travel demand~\cite{toole2015path} and the trips chains (a sequence of visited places with timestamps) of the population~\cite{schneider2013unravelling, jiang2016timegeo, xu2018planning}. However, methods solely relying on CDRs can not infer the routes taken by individuals due to coarse spatial resolution. As one kind of location-based services data, the geo-tagged check-in data provide more accurate locations than CDRs, but the data can be collected only when the users actively ``checked-in'' at their places of interest, making it a too infrequent source. Such data is activity-dependent and can not continuously record the traces of users~\cite{rashidi2017exploring}. 
Floating car data (FCD) are collected by transportation agencies for some specific purposes, recording the locations and speeds of floating cars. The GPS trajectories of taxis are the most commonly used FCD to analyze the traffic states. However, because of the cruising behavior of taxi drivers to search potential passengers, the taxis' trajectories are not perfect to study the route choice behavior of residents.
In this work, we explore the location-based services (LBS) data, which specifically refer to the collection of the check-in or trajectory data generated by a set of smartphone applications.
LBS use a smartphone's localization technology (i.e. GPS, Wi-Fi) to track the holder's location down to a street address, if the holder has opted-in to allow the service to do that. Compared to  the check-in data from one single application, our LBS data collect the locations from multiple applications and have much higher sampling frequency for two reasons, (i) some applications continuously collect the locations for providing map-related services to users; (ii) the aggregation of records from multiple applications also increase the sampling frequency. In recent years, LBS data have been used to examining meaningful visited places and social mixing~\cite{scherrer2018travelers,dong2019segregated}, travel behavior mining~\cite{liao2019individual}, and commuting pattern estimation~\cite{mcneill2017estimating}. The emergence of data collaborative using LBS records in the light of COVID19 pandemic has accelerated their use and value~\cite{aleta2020modeling,klein2020assessing}.

This paper aims at analyzing the LBS for urban scale mobility demand, with focus in gaining insights on their use to extract routing behavior. Focusing on the Dallas-Fort Worth (DFW) metroplex, we first describe the process to impute vehicular trips from LBS data and then present a framework to deal with sparse data. Utilizing LBS data to analyze routing behavior faces important challenges: (i) LBS data are collected when the applications are activated, whenever the users are staying in one place or moving in unknown travel modes; (ii) LBS data are collected with varying sample rates, which hinder the detection of actual routes. We present detailed steps to resolve these issues. Further, we analyze the change of routing behavior by connecting it with the number of trips, the travel distance, and the travel time during peak hours. 
The main contributions of this work are summarized as follows: (i) we present detailed steps to process the LBS data, extract the vehicular trios and detect routes without the use of the road network and map-matching; (ii) we analyze differences of travelers' routing behavior by different number of trips, travel distances, and the time periods in one day. (iii) we inspect the impact of traffic congestion on individuals' routing behavior using two metrics, the travel time index and buffer index. Empirical results confirm that individuals that explore more routes can reduce the impact of congestion and increase their reliability of travel times. The complete implementation of all of our data analysis framework can be found at \url{https://github.com/humnetlab/RoutingBehavior}.

The rest of the paper is organized as follows. In Sec. 2, we give an overview of the LBS data used in this work; Sec. 3 depicts the methodology to process the LBS data and find the travelers' routes; Sec. 4 analyzes the route choice behavior and its connection to multiple factors. Finally, we conclude the work in Sec. 5.

\section{Data Description}
\label{sec:data}

LBS are services offered to users through applications installed on smart mobile devices. Geographical locations of the users are simultaneously and actively collected by the application developers or map service operators. The users are normally positioned by global positioning system (GPS) or Wi-Fi positioning system (WPS), which are fairly accurate in space and offer new opportunities to study human activity and its complex interaction with the built environment at fine scale~\cite{kwan2004gis, ratti2006mobile, zheng2015trajectory, miller2015data, silva2019urban}.

The LBS data used in this work are provided by Cuebiq, a location intelligence and measurement platform~\cite{cuebiq2019}. The datasets cover the DFW metroplex in Texas and were collected over a period of 6 months, from November 1st, 2016 to April 30th, 2017. The total number of users is approximate to $6.5$ million and these users generated about $12.43$ billion records in the given region and time period. Each LBS record consists of the pseudonymized user ID, timestamp and geographical coordinate. Fig.~\ref{fig:data}(a) illustrates the covered region and the visitation count in each grid cell in the first week of November 2016. The entire region is divided into $512$$\times$$512$ cells with approximate size $360$$\times$$320$~m$^2$, down to the block level. The highlighting of freeways and downtown in the heatmap indicates that they are the busiest places in terms of visitation counts.


As LBS data are collected when the user is interacting with the application, the collection can be interrupted if the user stops using the application. Besides, the applications are used with variant frequency. Thus, the users in LBS data have different numbers of records and timespan, which is defined as the time difference between the first and last records of the user. In Fig.~\ref{fig:data}(b), we show the timespan versus the number of records for all users in a heatmap. The region with dark green indicates that a large number of users are associated with the corresponding timespan and the number of records. As we can observe, a considerable part of users were recorded during a long term but have small numbers of records as they are not using the applications frequently. Other users have a considerable number of records but short timespan. These might be temporary visitors to the DFW metroplex or short time adopters of the app. For exploring the routing behavior, we require long-term observation of moving traces. To that end, we select the users whose data are collected over $60$ days or more and have more than $1,000$ records, as enclosed by a red rectangle in Fig.~\ref{fig:data}(b). As a result, $13\%$ of the users and $86\%$ of the records are kept for further analysis.

An important challenge remains, even with this sample, the records of LBS data are collected with variant, usually low frequency, because of the intermittent use of applications and different sample rates of applications. The LBS datasets are collected by a number of mobile applications (Apps) when the mobile phone user is interacting with these Apps or keeps these Apps running in the background. The sample interval, defined as the time difference between two consecutive records of the same user, is not fixed. Fig.~\ref{fig:data}(c) shows the distribution of sample intervals in the LBS data. The sample intervals of a large proportion of records are larger than 2 min, which much lower frequency than floating car data and hinder the routing behavior analysis, especially in dense road networks. Besides, the uneven sample rates shown in Fig. 1(c) are mainly caused by the aggregation of records from multiple applications. Next, we propose a method to deal with this limitation and extract some valuable information from this data source.

\section{Methodology}
\label{sec:methods}

For analyzing the route choice behavior, a primary task is to map the users to specific routes they were taking. However, tracking the routes from LBS data is a challenge in the following two aspects: (i) LBS collect the data of users when they stay or move with all kinds of travel modes, e.g., walking, biking, driving and public transportation. Vehicular trips must first be imputed from the raw data for further route choice behavior analysis; (ii) LBS data are collected from multiple applications with different sample rates and at low-resolution. This hinders the entire routes over the road networks. Once the vehicular trips are assigned, we then select high-resolution trips for route detection and find the routes of other low-resolution trips by aligning them with the high-resolution ones.

\subsection{Vehicular Trips Detection}
\label{sec:vehTrip}

For the records of each user, we first partition her records into a sequence of trips by looking into the time difference between two consecutive records. After the user selection illustrated in Fig.~\ref{fig:data}(b), the remaining users are labeled as high-frequency ones, a.k.a, active users. In this context, we suppose that a user starts a new trip if there is no record for at least 30 minutes before the current one, that is, $t_{current} - t_{previous} \geq 30\ min$. Then we drop out the trips with less than 5 records. At this point, the records of each user have been partitioned into a sequence of trips in all kinds of travel modes. 

A number of methods have been proposed to derive the travel mode from trajectory data, most of them process the high-resolution GPS traces or utilize sophisticated learning methods that require gold labels of travel modes for model training~\cite{xiao2015travel, dabiri2018inferring}. Here we use a simple rule to identify the trip as a vehicular trip if its average speed is between $20$ km/h and $100$ km/h, leaving room for further improvements. In addition, there are trips with outliers caused by the GPS drift, which are eliminated in our experiments. We label the points which have less than 50 neighbors in the set of points of all vehicular trips within 100 m radius as outliers. The entire vehicular trip is rejected once there are a considerable proportion of points in a vehicular trip (i.e., more than $20\%$) labeled as outliers. This method might also remove the trips taking the routes which are rarely used. But it would not impact our analysis as we place emphasis on the commonly used routes by each user. The pseudocode for deriving the vehicular trips from the raw LBS data is depicted in Algorithm~\ref{alg:vehicleTrips}.

\makeatletter
    \def\BState{\State\hskip-\ALG@thistlm}
    \makeatother
    \begin{algorithm}[htb!]
        \caption{Vehicular Trips Deriving}\label{alg:vehicleTrips}
        \begin{algorithmic}[1]
            \State \textbf{Step 1: User selection} 
            \For{$ user \in Users$}  \Comment{Loops through each user in the raw data}
                \State Count the timespan and the number of records
                \If {$timespan < 60 \ days$ or $\#\ records < 1,000,$}
                \State Remove the user's records
                \EndIf
            \EndFor
            \State \textbf{return} selected data \vspace{2mm}
            \State \textbf{Step 2: Vehicular trips detection}
            \For{$ user \in Users$}  \Comment{Loops through each user in the selected data}
                \For{$item \in records$} \Comment{Loop through each item in user's records}
                    \State Calculate the time difference between the current and next item, $\Delta t$
                    \If {$\Delta t \geq 30 \ min$}
                    \State Partition the trace
                    \EndIf
                \EndFor
                \State \textbf{User's trips are ready.}
                \For{$trip \in Trips$} \Comment{Loop through each trip in user's trips}
                    \State Calculate the average travel speed, $\bar v$
                    \If {$\bar v < 20 \ km/h$ or $\bar v > 100 \ km/h$}
                    \State Remove the trip
                    \EndIf
                    \State Update $trip Id$
                \EndFor
            \EndFor
            \State \textbf{return} vehicular trips \vspace{2mm}
            \State \textbf{Step 3: Vehicular trips filtering}
            \For{$ user \in Users$}  \Comment{Loops through each user with vehicular trips}
                \For{$trip \in Trips$} \Comment{Loop through each trip in user's trips}
                    \State Number of removed points, $N_{outlier} = 0$
                    \For {$point \in trip$}
                    \State Count neighboring points within $100m$ in the data of all selected users, $N_{neighbor}$
                    \If { $N_{neighbor} < 50$}
                    \State $N_{outlier} \mathrel{+}= 1$
                    \EndIf
                    \EndFor
                    \If { $N_{outlier} / length(trip) \geq 0.2$}
                    \State Remove the trip
                    \EndIf
                    \State Update $trip Id$
                \EndFor
            \EndFor
            \State \textbf{return} selected vehicular trips
        \end{algorithmic}
    \end{algorithm}

\subsection{Collective Travel Demand Estimation and Validation}

Similar to travel behavior analysis using CDR data~\cite{jiang2013review, toole2015path, ccolak2015analyzing}, we first detect the possible home locations of each active LBS user. To this end, we collect the stay locations for each user from the origins and destinations of all vehicular trips. Each stay location is associated with the departure or arrival time. As the users usually depart from home in the morning and arrive home in the end of the day, we select all departure locations between 5:00 a.m. and 10:00 a.m. and the arrival locations after 5:00 p.m. every day to compose the user's home candidate pool. If more than 30 locations are found, we then cluster the candidate locations using DBSCAN, setting the spatial threshold as 300 m, considering the users may park their vehicles around the significant places. We define the centroid of the largest cluster as a home place if the fraction of points in this cluster is larger than $40\%$. In Fig.~\ref{fig:home}(a), we show the detected home locations of the active LBS users. The accuracy of home detection is always challenging due to the lack of ground truth. Vanhoof et al. used the census data to validate the home location detection methods~\cite{vanhoof2018assessing}. However, the validation can not be very reliable even at collective level because of the heterogeneous distribution of active LBS users in space. Note that we use the LBS users’ home locations at ZIP code level to expand the users' travel demand, suggesting that we do not need to identify the home location within a few meters. We compare the active LBS users settling in each ZIP code versus its population from the U.S. census data~\cite{census2016}, and find positive Pearson correlation ($\rho=0.73$), as shown in Fig.~\ref{fig:home}(b). However, there do exist some ZIP codes with large populations but very small numbers of active LBS users, indicating the difficulty of user expansion.


Besides, we aggregate the derived vehicular trips by an hour to achieve the hourly flow during one week, as shown in Fig.~\ref{fig:demand}(a), displaying morning and evening peaks on weekdays and midday peaks on weekends. We next inspect the vehicle trips from LBS data by comparing the origin-destination (OD) flow with the travel survey conducted by the North Central Texas Council of Governments (NCTCOG) in 2014~\cite{nctcog2014}. As the LBS data are collected from a fraction of the population, like the CDR data~\cite{toole2015path, ccolak2015analyzing}, we define the expansion factors in each ZIP code as the ratio of the population from 2016 U.S. census data to the number of LBS users living in the same region. The distribution of expansion factors is presented in Fig.~\ref{fig:demand}(b), and the 1st, 2nd, and 3rd quartiles of the expansion factors are 136.6, 217.2, and 356.8, respectively. Note that population synthesis is one more advanced way to expand the active users to population level than our expansion factor. The population synthesis expands the users with their demographic/socioeconomic information and sophisticated models~\cite{sun2015bayesian}.
The expansion factor of each Zip code is visualized in Fig.~\ref{fig:demand}(c). The Zip codes in the urban area generally have smaller expansion factors than rural area. We then aggregate the vehicular trips during the morning peak hours (6:30 - 9:00 a.m.) at Zip code level and scale the flow with the expansion factors. In Fig.~\ref{fig:demand}(d), we compare the values of vehicular travel demand for all OD pairs at ZIP code level between the expanded LBS flow and the NCTCOG survey in the morning peak hours, and find the linear fitting slop equals to $0.82$ and $r^2 = 0.59$. Fig.~\ref{fig:demand}(e) illustrates the spatial distribution of vehicular travel flow above 0.01\% of the total demand during the morning peak hours achieved from the expanded LBS data and NCTCOG data, respectively. Even we show the estimated travel demand is visually comparable to the NCTCOG survey data, the Pearson correlation only reaches $0.79$. That can be caused by several reasons in this work, such as (i) we simply selected active users with the timespan and number of records in the raw LBS data, aiming at removing the temporary visitors in the DFW metroplex. However, we can not accurately identify residents from all LBS users with such simple rules; (ii) The distribution of active LBS users is different to the residents in space, as shown in Fig. 2(b) and the spatial distribution of expansion factors in Fig. 3(c); (iii) As we used simple rules to identify the vehicular trips, some non-vehicular trips are kept in our OD matrix; (iv) we used simple expansion factors to expand the travel demand of active LBS users to the population.


\subsection{Route Detection}
\label{sec:route}

The core challenge of deriving route choice behavior from LBS data is the varying sample interval of the records, as shown in Fig.~\ref{fig:data}(c). The varying sample interval in time leads to the heterogeneity of displacement between two consecutive records, ranging from several meters to kilometers. Such heterogeneity would cause the incorrect calculation of the similarity between two trips, and affect the clustering of trips. For instance, even when two low-resolution trips are taking the same route, the similarity between them would be low as the distance between the point pair would be large. One of the popular solutions is to map the points to the road network with map-matching and connect the distant consecutive points with the shortest path in the road network. However, it requires the road network and map-matching is computationally expensive for massive trajectory data~\cite{chen2014map}.


To overcome this challenge, we design a simple yet efficient two-step procedure to find the routes: (i) selecting the high-resolution trips, in which the maximum distance gap between consecutive points is less than 1~$km$ and detect the taken routes by trace clustering; (ii) matching the low-resolution trips to the high-resolution ones and finding the most likely taken routes. Fig.~\ref{fig:routes}(a) presents the extracted vehicular trips from the raw LBS records for 200 sample users. The layout of vehicular trips displays a good match with the road networks in the DFW metroplex. We select one user to illustrate the two-step procedure, as shown in Fig.~\ref{fig:routes}(b). For understanding the route choice behavior, we decide to focus on the frequently visited places between which there are repeated number of trips. To this end, we cluster the origins and destinations of all trips using DBSCAN and label the centroids of clusters as significant places. For each active LBS user, we then select two unidirectional OD pairs for further route detection, the OD pairs with the largest and the second largest numbers of trips, as illustrated in Fig.~\ref{fig:routes}(c). Note that the two selected OD pairs may not be reversed. After this step, we keep 1,194,154 trips ($5.3\%$ of all trips after trip segmentation) of 58,333 users ($0.9\%$ of all users in the raw LBS data) for routing behavior analysis.

Among the trips between a selected origin and destination pair, we first select the high-resolution trips to label the routes. This is because the inference is more reliable when the distance gaps between consecutive points are small. The high-resolution trips are grouped to one or more clusters using a clustering method described in the following if there is more than one trip. The purpose of trip clustering is to group the trips which are taking the same route. There are two selection criteria for trip clustering, measurement of trip similarity and the number of clusters. Two of the most popular measurements are the longest common subsequence (LCSS) and dynamic time warping (DTW)~\cite{kim2015spatial, zheng2015trajectory}. However, Atev et al. proposed a modified Hausdorff distance and confirmed that it could surpass both LCSS and DTW in trajectory clustering~\cite{atev2010clustering}. In fact, we find that the modified Hausdorff improves its robustness to the noise by rejecting a number of worst matches of points in the two trajectories. In this work, we adopt the same modified Hausdorff to calculate the distance between two high-resolution trips and DBSCAN to cluster them into one or more groups. Ideally, each cluster represents one route. For the sample user in Fig.~\ref{fig:routes}(b), there are three routes detected on the high-resolution trips, differentiated by color in Fig.~\ref{fig:routes}(d).

In the second step, we add the records of the low-resolution trips by aligning them with the high-resolution ones. Specifically, for each low-resolution trip, we first calculate the maximum distance between the point sets in it and the sets in each high-resolution trip. This distance indicates how far does this trip deviate from the high-resolution cluster and is used to decide if they belong to the same route. If the target low-resolution trip has a nearest high-resolution trip within a certain distance (e.g., 1 km), we identify its route the same as the high-resolution one. Otherwise, we remove this low-resolution trip as its route is uncertain. Fig.~\ref{fig:routes}(e) presents the final route detection results for the sample user. The detailed pseudocode for route detection is depicted in Algorithm~\ref{alg:routes}.

\makeatletter
    \def\BState{\State\hskip-\ALG@thistlm}
    \makeatother
    \begin{algorithm}[htb!]
        \caption{Route Detection}\label{alg:routes}
        \begin{algorithmic}[1]
            \State \textbf{Step 1: Top OD pair selection}
            \State \textbf{1.1:} Find stay locations via DBSCAN clustering the origins and destinations
                \If {$\#\ points\ in\ cluster \geq 5$,}
                \State Keep the centroid of cluster as a stay location
                \EndIf
            \State \textbf{1.2:} Count the number of trips between stay locations
            \State \textbf{return} OD pairs with the largest and the second largest number of trips
            \vspace{2mm}
            \State \textbf{Step 2: Route detection on high-resolution trips}
            \State \textbf{2.1:} Finding high-resolution trips
            \For{$ trip \in Top~OD~pair$}  \Comment{Loops through each trip in the selected trips}
            \State Calculate the distance gap between consecutive points, find the maximum gap $g_{max}$
                \If {$g_{max} \leq 1\ km$}
                \State Label the trip as high resolution
                \EndIf
            \EndFor
            \State \textbf{2.2:} Do trajectory clustering on all high-resolution trips using DBSCAN
            \State \textbf{2.3:} Label each cluster as one route
            \State \textbf{return} high-resolution trips with labeled routes
            \vspace{2mm}
            \State \textbf{Step 3: Route labeling for low-resolution trips}
            \For{$ trip_L \in low\mbox{-}resolution\ trips$}  \Comment{Loops through each trip in the low-resolution trips}
                \For{$trip_H \in high\mbox{-}resolution\ trips$} 
                    \State Calculate the maximum distance from points in $trip_L$ to $trip_H$
                \EndFor
            \State Find the nearest $trip_{H,nearest}$ and the distance $D_{H,nearest}$
            \If {$D_{H,nearest} \leq 1\ km$}
            \State Give the route label of $trip_{H,nearest}$ to $trip_L$
            \Else
            \State Route taken by $trip_L$ is uncertain, remove it
            \EndIf
            \EndFor
            \State \textbf{return} trips with labeled routes
        \end{algorithmic}
    \end{algorithm}

\section{Route Choice Behavior Analysis}
\label{sec:analysis}
\subsection{Distribution of number of Routes}


We first inspect the statistical distributions of the number of trips $N_{trip}$ and the number of routes $N_{route}$ in the selected top two OD pairs for all active users. Fig.~\ref{fig:distribution}(a) and (b) present the distributions of $N_{trip}$ and $N_{route}$, respectively. Log-normal distributions resemble the data in both cases, in agreement with Lima et al.'s findings~\cite{lima2016understanding}.
The mean value of $N_{trip}$ reaches $29.08$, while the mean value of the $N_{route}$ between these OD pairs is $1.56$. From Fig.~\ref{fig:distribution}(b), we observe that among all active users in our LBS data, $51.35\%$ of them only take one route to complete the top OD pairs; $37.5\%$ of them take two routes and only $11.15\%$ of them take more than 2 routes.

Next, we inspect the discrepancy of routing behavior during peak and off-peak hours. To this end, we split all trips in users' top OD pairs into four groups by their departure time, e.g., morning peak hours from 7:00 to 10:00 (AM), midday from 10:00 to 16:00 (MD), evening peak hours from 16:00 to 19:00 (PM) and the rest of the day (RD). We then count $N_{route}$ of each user in these four time periods on weekdays and weekends, respectively. Fig.~\ref{fig:distribution}(c) presents the fraction of active users taking different $N_{route}$ during each time period on weekdays. The number of trips in each time period is presented in the inset. We observe the fractions of users taking 2 routes and above during AM and PM are apparently higher than the other two periods, suggesting that the users tend to take more routes during the peak hours to finish their trips more efficiently, e.g., in shorter travel time. As for the routing behavior on weekends shown in Fig.~\ref{fig:distribution}(d), the distribution of $N_{route}$ changes little between time periods due to the traffic on weekends is not as congested as weekdays.

\subsection{Route Choice Behavior of Different Groups of Travelers}


For comprehensive understanding of the discrepancy of route choice behavior among the travelers, we group them by their travel frequencies and travel distances. According to the distribution of $N_{trip}$ presented in Fig.~\ref{fig:distribution}(a), we split the travelers into four groups according to the following rules, $N_{trip} < 20$, $20 \leq N_{trip} < 40$, $40 \leq N_{trip} < 60$, and $N_{trip} \geq 60$. Fig.~\ref{fig:groups}(a) presents the distribution of $N_{route}$ per group. We notice that frequent travelers tend to explore more routes than non-frequent travelers, most likely because the frequent travelers know the traffic congestion better and are more confident to find efficient alternatives. The phenomenon also can be confirmed in Fig.~\ref{fig:groups}(b), which presents the distribution of $N_{trip}$ of users taking different number of routes in their routine OD pairs. The median value of $N_{trip}$ of travelers with more than three routes is around $60$, while the median value of travelers sticking on one route is less than $30$.

The distance between origin and destination could be one of the factors that affect the number of routes selected as the distance determines the number of candidate routes in a given road network. We then compare the routing behavior of travelers with different ranges of travel distance. The travelers are grouped into Q1 to Q4 by the $25$th, $50$th, $75$th percentiles of their travel displacements. Fig.~\ref{fig:groups}(c) depicts the distribution of $N_{route}$ of each group, and Fig.~\ref{fig:groups}(d) depicts the distribution of travel displacements of travelers with different $N_{route}$. As expected, we can observe that most of the users with short trips in Q1 stick on only one route. From Fig.~\ref{fig:groups}(d), we can see the peak of the distribution is around 5~km for the users who only take one route; while the peak is around 10~km for the users who take more than 3 routes. These observations indicate that more routes are likely been taken if the users make longer trips between two significant places. It can be explained from the perspective of network. In a dense road network, the larger the distance between two nodes is, the more alternative routes with similar cost the travelers can choose.

\subsection{Route Choice Behavior in Traffic Congestion}

Traffic congestion is a major consideration driving travelers to find alternatives, especially during peak hours. Here, we investigate the relation between travel time and the number of routes in the routine OD pairs of active travelers. For each user, we calculate the travel time index (TTI) of all trips in travelers' top OD pairs to assess the additional travel time caused by congestion. Given a number of trips in an OD pair, TTI is defined as the ratio of the average travel time to the free flow travel time, 
\begin{equation}
    TTI = T_{avg} / T_{free} \ ,
\end{equation}
where $T_{avg}$ refers to the average travel time of all trips made by one user; $T_{free}$ refers to the free flow travel time from her origin to destination, approximated by the minimum travel time among all trips in the OD pair. The larger the TTI is, the more traffic congestion the user met during the routine journey. The introducing of TTI enables us to compare the travel delay of OD pairs even they have various travel distances. Fig.~\ref{fig:traveltime}(a) and (b) illustrate the TTI of the travelers with different $N_{route}$ during the AM (7:00-10:00) and PM (16:00-19:00) peak hours on weekdays, respectively. It is noticeable that travelers taking more routes tend to have lower TTI. The average TTI values are also illustrated in Table~\ref{tab:traveltime}. To reduce the impact of the extreme small or large values of TTI, we also present the mean value of the TTI falls into the $95\%$ confidence interval and the standard deviation (STD) in Table~\ref{tab:traveltime}. We can conclude that travelers with flexible route choice behavior can lower their travel time by avoiding traffic congestion in the primary routes. The insets of Fig.~\ref{fig:traveltime}(a) and (b) present the distribution of TTI for all travelers, showing the average TTI is nearly $2.0$ during the peak hours. This reveals that, because of the congestion, the travelers in DFW metroplex spent nearly double free flow travel time to complete their journeys during rush hours.



Beyond the additional traffic time caused by congestion, the reliability of travel time is also significant to many travelers, especially when they need to arrive at the destination on time. Reliability has been considered as a key performance measure by transportation planners and decision-makers. We introduce the buffer index (BI) to assess the travel time reliability of all trips in the traveler's top OD pair~\cite{TTRreport}. The BI represents the extra buffer time that the traveler should add to the average travel time when planning trips to ensure on-time arrival. Here we define BI as the relative gap between the $85$th percentile travel time and the average travel time of an OD pair, 
\begin{equation}
    BI = (T_{85th} - T_{avg}) / T_{avg} \times 100\% \ .
\end{equation}
The BI is expressed as a percentage and its value increases as reliability gets worse. Fig.~\ref{fig:traveltime}(c) and (d) present the BI of the travelers with different $N_{route}$ during the AM and PM peak hours on weekdays, respectively. The average BI values, the average BI in the $95\%$ confidence interval, and the STD of BI are illustrated in Table~\ref{tab:traveltime}. We notice that the average BI decreases along with the increase of $N_{route}$, suggesting that the travelers are changing their routes considering the real-time traffic to increase the reliability of travel time.

\section{Conclusion and outlook}
Understanding the route choice behavior is an essential task for not only modeling human mobility in transportation networks but also route management to relieve traffic congestion. In this paper, we presented a data analysis framework for understanding route choice behavior with massive LBS data. Steps include, user selection, vehicular trip enrichment, trip clustering, route detection, and behavior analysis. We analyzed the six-month LBS data in the Dallas-Fort Worth metroplex, and selected the trips between the most frequent origin-destination pair of each user for understanding routing behavior. We found that the distribution of the number of routes can be modeled by a log-normal distribution. We also inspected the relation between the number of routes and the travel displacement and found that travelers with longer travel distances tend to select more routes to shorten their travel time. 
We also confirmed that travelers take more routes during peak hours than off-peak hours, and those individuals that explore more routes reduce their impact of congestion and increase their reliability of travel times. The proposed framework makes LBS data useful to evaluate the route choice behavior of different groups of travelers and their reaction to traffic congestion. As future applications, this could be implemented to evaluate traffic regulation strategies, such as the congestion charges.

Moreover, there are still some directions for further study. For instance, (i) for comprehensively understanding people's emphasis on different travel costs (i.e., travel time, routing distance, etc.), we need to further integrate these factors per route per user. To that end, we need to estimate the traffic states in the entire road network through map-matching; (ii) this work presented a case study in a region without congestion pricing. The relation between socio-economic characteristics and routing behavior merits more attention in cities with traffic interventions. 



\begin{backmatter}

\section*{Availability of data and material}
The complete implementation of all of our data analysis framework can be found at \url{https://github.com/humnetlab/RoutingBehavior}. Mobility data are provided by Cuebiq, a location intelligence and measurement platform. Through its Data for Good program (https://www.cuebiq.com/about/data-for-good/), Cuebiq provides access to aggregated and privacy-enhanced mobility (see below) data for academic research and humanitarian initiatives. These first-party data are collected from users who have opted in to provide access to their GPS location data and their ids are pseudonymized. 
In order to preserve privacy, noise is added to these ``personal areas”, by up-leveling these areas to the Census Block Group Level. This allows for demographic analysis while obfuscating the true home location of anonymous users and prohibiting misuse of data.

\section*{Competing interests}
The authors declare that they have no competing interests.

\section*{Funding}
This work was supported by the MIT Energy Initiative and the Berkeley Deep Drive consortium.

\section*{Author's contributions}
Y.X. and M.C.G. conceived the research and designed the analyses. Y.X. and R.D.C. processed and analyzed the data. Y.X. and M.C.G. performed the results analyses and wrote the paper. M.C.G. provided general advice and supervised the research.

\section*{Acknowledgements}
Authors thank Ricardo Sanchez Gomez and his team at Cintra/Ferrovial to initiating the case study that motivated this work, and Antonio Lima for inspiring some of the findings in this topic.


\bibliographystyle{bmc-mathphys} 
\bibliography{bmc_article}      


\begin{thebibliography}{47}
\ifx \bisbn   \undefined \def \bisbn  #1{ISBN #1}\fi
\ifx \binits  \undefined \def \binits#1{#1}\fi
\ifx \bauthor  \undefined \def \bauthor#1{#1}\fi
\ifx \batitle  \undefined \def \batitle#1{#1}\fi
\ifx \bjtitle  \undefined \def \bjtitle#1{#1}\fi
\ifx \bvolume  \undefined \def \bvolume#1{\textbf{#1}}\fi
\ifx \byear  \undefined \def \byear#1{#1}\fi
\ifx \bissue  \undefined \def \bissue#1{#1}\fi
\ifx \bfpage  \undefined \def \bfpage#1{#1}\fi
\ifx \blpage  \undefined \def \blpage #1{#1}\fi
\ifx \burl  \undefined \def \burl#1{\textsf{#1}}\fi
\ifx \doiurl  \undefined \def \doiurl#1{\textsf{#1}}\fi
\ifx \betal  \undefined \def \betal{\textit{et al.}}\fi
\ifx \binstitute  \undefined \def \binstitute#1{#1}\fi
\ifx \binstitutionaled  \undefined \def \binstitutionaled#1{#1}\fi
\ifx \bctitle  \undefined \def \bctitle#1{#1}\fi
\ifx \beditor  \undefined \def \beditor#1{#1}\fi
\ifx \bpublisher  \undefined \def \bpublisher#1{#1}\fi
\ifx \bbtitle  \undefined \def \bbtitle#1{#1}\fi
\ifx \bedition  \undefined \def \bedition#1{#1}\fi
\ifx \bseriesno  \undefined \def \bseriesno#1{#1}\fi
\ifx \blocation  \undefined \def \blocation#1{#1}\fi
\ifx \bsertitle  \undefined \def \bsertitle#1{#1}\fi
\ifx \bsnm \undefined \def \bsnm#1{#1}\fi
\ifx \bsuffix \undefined \def \bsuffix#1{#1}\fi
\ifx \bparticle \undefined \def \bparticle#1{#1}\fi
\ifx \barticle \undefined \def \barticle#1{#1}\fi
\ifx \bconfdate \undefined \def \bconfdate #1{#1}\fi
\ifx \botherref \undefined \def \botherref #1{#1}\fi
\ifx \url \undefined \def \url#1{\textsf{#1}}\fi
\ifx \bchapter \undefined \def \bchapter#1{#1}\fi
\ifx \bbook \undefined \def \bbook#1{#1}\fi
\ifx \bcomment \undefined \def \bcomment#1{#1}\fi
\ifx \oauthor \undefined \def \oauthor#1{#1}\fi
\ifx \citeauthoryear \undefined \def \citeauthoryear#1{#1}\fi
\ifx \endbibitem  \undefined \def \endbibitem {}\fi
\ifx \bconflocation  \undefined \def \bconflocation#1{#1}\fi
\ifx \arxivurl  \undefined \def \arxivurl#1{\textsf{#1}}\fi
\csname PreBibitemsHook\endcsname

\bibitem{xu2020deconstructing}
\begin{barticle}
\bauthor{\bsnm{Xu}, \binits{Y.}},
\bauthor{\bsnm{Olmos}, \binits{L.E.}},
\bauthor{\bsnm{Abbar}, \binits{S.}},
\bauthor{\bsnm{Gonz{\'a}lez}, \binits{M.C.}}:
\batitle{Deconstructing laws of accessibility and facility distribution in
  cities}.
\bjtitle{Science advances}
\bvolume{6}(\bissue{37}),
\bfpage{4112}
(\byear{2020})
\end{barticle}
\endbibitem

\bibitem{weisbrod2003measuring}
\begin{botherref}
\oauthor{\bsnm{Weisbrod}, \binits{G.}},
\oauthor{\bsnm{Vary}, \binits{D.}},
\oauthor{\bsnm{Treyz}, \binits{G.}}:
Measuring economic costs of urban traffic congestion to business.
Transportation Research Record: Journal of the Transportation Research Board
(1839),
98--106
(2003)
\end{botherref}
\endbibitem

\bibitem{jiang2017transport}
\begin{barticle}
\bauthor{\bsnm{Jiang}, \binits{B.}},
\bauthor{\bsnm{Liang}, \binits{S.}},
\bauthor{\bsnm{Peng}, \binits{Z.-R.}},
\bauthor{\bsnm{Cong}, \binits{H.}},
\bauthor{\bsnm{Levy}, \binits{M.}},
\bauthor{\bsnm{Cheng}, \binits{Q.}},
\bauthor{\bsnm{Wang}, \binits{T.}},
\bauthor{\bsnm{Remais}, \binits{J.V.}}:
\batitle{{Transport and public health in China: the road to a healthy future}}.
\bjtitle{The Lancet}
\bvolume{390}(\bissue{10104}),
\bfpage{1781}--\blpage{1791}
(\byear{2017})
\end{barticle}
\endbibitem

\bibitem{xu2019unraveling}
\begin{barticle}
\bauthor{\bsnm{Xu}, \binits{Y.}},
\bauthor{\bsnm{Jiang}, \binits{S.}},
\bauthor{\bsnm{Li}, \binits{R.}},
\bauthor{\bsnm{Zhang}, \binits{J.}},
\bauthor{\bsnm{Zhao}, \binits{J.}},
\bauthor{\bsnm{Abbar}, \binits{S.}},
\bauthor{\bsnm{Gonz{\'a}lez}, \binits{M.C.}}:
\batitle{{Unraveling environmental justice in ambient PM$_{2.5}$ exposure in
  Beijing: A big data approach}}.
\bjtitle{Computers, Environment and Urban Systems}
\bvolume{75},
\bfpage{12}--\blpage{21}
(\byear{2019})
\end{barticle}
\endbibitem

\bibitem{hu1997day}
\begin{barticle}
\bauthor{\bsnm{Hu}, \binits{T.-Y.}},
\bauthor{\bsnm{Mahmassani}, \binits{H.S.}}:
\batitle{Day-to-day evolution of network flows under real-time information and
  reactive signal control}.
\bjtitle{Transportation Research Part C: Emerging Technologies}
\bvolume{5}(\bissue{1}),
\bfpage{51}--\blpage{69}
(\byear{1997})
\end{barticle}
\endbibitem

\bibitem{kelly2016transport}
\begin{barticle}
\bauthor{\bsnm{Kelly}, \binits{F.J.}},
\bauthor{\bsnm{Zhu}, \binits{T.}}:
\batitle{Transport solutions for cleaner air}.
\bjtitle{Science}
\bvolume{352}(\bissue{6288}),
\bfpage{934}--\blpage{936}
(\byear{2016})
\end{barticle}
\endbibitem

\bibitem{xu2017collective}
\begin{barticle}
\bauthor{\bsnm{Xu}, \binits{Y.}},
\bauthor{\bsnm{Gonz{\'a}lez}, \binits{M.C.}}:
\batitle{Collective benefits in traffic during mega events via the use of
  information technologies}.
\bjtitle{Journal of The Royal Society Interface}
\bvolume{14}(\bissue{129}),
\bfpage{20161041}
(\byear{2017})
\end{barticle}
\endbibitem

\bibitem{ccolak2016understanding}
\begin{barticle}
\bauthor{\bsnm{{\c{C}}olak}, \binits{S.}},
\bauthor{\bsnm{Lima}, \binits{A.}},
\bauthor{\bsnm{Gonz{\'a}lez}, \binits{M.C.}}:
\batitle{Understanding congested travel in urban areas}.
\bjtitle{Nature communications}
\bvolume{7}(\bissue{1}),
\bfpage{1}--\blpage{8}
(\byear{2016})
\end{barticle}
\endbibitem

\bibitem{prud2005london}
\begin{barticle}
\bauthor{\bsnm{Prud'homme}, \binits{R.}},
\bauthor{\bsnm{Bocarejo}, \binits{J.P.}}:
\batitle{The london congestion charge: a tentative economic appraisal}.
\bjtitle{Transport Policy}
\bvolume{12}(\bissue{3}),
\bfpage{279}--\blpage{287}
(\byear{2005})
\end{barticle}
\endbibitem

\bibitem{wu2018stabilizing}
\begin{bchapter}
\bauthor{\bsnm{Wu}, \binits{C.}},
\bauthor{\bsnm{Bayen}, \binits{A.M.}},
\bauthor{\bsnm{Mehta}, \binits{A.}}:
\bctitle{Stabilizing traffic with autonomous vehicles}.
In: \bbtitle{2018 IEEE International Conference on Robotics and Automation
  (ICRA)},
pp. \bfpage{1}--\blpage{7}
(\byear{2018}).
\bcomment{IEEE}
\end{bchapter}
\endbibitem

\bibitem{gonzalez2008understanding}
\begin{barticle}
\bauthor{\bsnm{Gonzalez}, \binits{M.C.}},
\bauthor{\bsnm{Hidalgo}, \binits{C.A.}},
\bauthor{\bsnm{Barabasi}, \binits{A.-L.}}:
\batitle{Understanding individual human mobility patterns}.
\bjtitle{Nature}
\bvolume{453}(\bissue{7196}),
\bfpage{779}
(\byear{2008})
\end{barticle}
\endbibitem

\bibitem{pappalardo2015returners}
\begin{barticle}
\bauthor{\bsnm{Pappalardo}, \binits{L.}},
\bauthor{\bsnm{Simini}, \binits{F.}},
\bauthor{\bsnm{Rinzivillo}, \binits{S.}},
\bauthor{\bsnm{Pedreschi}, \binits{D.}},
\bauthor{\bsnm{Giannotti}, \binits{F.}},
\bauthor{\bsnm{Barab{\'a}si}, \binits{A.-L.}}:
\batitle{Returners and explorers dichotomy in human mobility}.
\bjtitle{Nature communications}
\bvolume{6},
\bfpage{8166}
(\byear{2015})
\end{barticle}
\endbibitem

\bibitem{jiang2016timegeo}
\begin{barticle}
\bauthor{\bsnm{Jiang}, \binits{S.}},
\bauthor{\bsnm{Yang}, \binits{Y.}},
\bauthor{\bsnm{Gupta}, \binits{S.}},
\bauthor{\bsnm{Veneziano}, \binits{D.}},
\bauthor{\bsnm{Athavale}, \binits{S.}},
\bauthor{\bsnm{Gonz{\'a}lez}, \binits{M.C.}}:
\batitle{{The TimeGeo modeling framework for urban mobility without travel
  surveys}}.
\bjtitle{Proceedings of the National Academy of Sciences}
\bvolume{113}(\bissue{37}),
\bfpage{5370}--\blpage{5378}
(\byear{2016})
\end{barticle}
\endbibitem

\bibitem{ben1999discrete}
\begin{bchapter}
\bauthor{\bsnm{Ben-Akiva}, \binits{M.}},
\bauthor{\bsnm{Bierlaire}, \binits{M.}}:
\bctitle{Discrete choice methods and their applications to short term travel
  decisions}.
In: \bbtitle{Handbook of Transportation Science},
pp. \bfpage{5}--\blpage{33}.
\bpublisher{Springer},
\blocation{Boston, MA}
(\byear{1999})
\end{bchapter}
\endbibitem

\bibitem{prato2009route}
\begin{barticle}
\bauthor{\bsnm{Prato}, \binits{C.G.}}:
\batitle{Route choice modeling: past, present and future research directions}.
\bjtitle{Journal of choice modelling}
\bvolume{2}(\bissue{1}),
\bfpage{65}--\blpage{100}
(\byear{2009})
\end{barticle}
\endbibitem

\bibitem{di2016boundedly}
\begin{barticle}
\bauthor{\bsnm{Di}, \binits{X.}},
\bauthor{\bsnm{Liu}, \binits{H.X.}}:
\batitle{Boundedly rational route choice behavior: A review of models and
  methodologies}.
\bjtitle{Transportation Research Part B: Methodological}
\bvolume{85},
\bfpage{142}--\blpage{179}
(\byear{2016})
\end{barticle}
\endbibitem

\bibitem{zhu2015people}
\begin{barticle}
\bauthor{\bsnm{Zhu}, \binits{S.}},
\bauthor{\bsnm{Levinson}, \binits{D.}}:
\batitle{Do people use the shortest path? an empirical test of wardrop’s
  first principle}.
\bjtitle{PloS one}
\bvolume{10}(\bissue{8}),
\bfpage{0134322}
(\byear{2015})
\end{barticle}
\endbibitem

\bibitem{lima2016understanding}
\begin{barticle}
\bauthor{\bsnm{Lima}, \binits{A.}},
\bauthor{\bsnm{Stanojevic}, \binits{R.}},
\bauthor{\bsnm{Papagiannaki}, \binits{D.}},
\bauthor{\bsnm{Rodriguez}, \binits{P.}},
\bauthor{\bsnm{Gonz{\'a}lez}, \binits{M.C.}}:
\batitle{Understanding individual routing behaviour}.
\bjtitle{Journal of The Royal Society Interface}
\bvolume{13}(\bissue{116}),
\bfpage{20160021}
(\byear{2016})
\end{barticle}
\endbibitem

\bibitem{di2018sequences}
\begin{barticle}
\bauthor{\bsnm{Di~Clemente}, \binits{R.}},
\bauthor{\bsnm{Luengo-Oroz}, \binits{M.}},
\bauthor{\bsnm{Travizano}, \binits{M.}},
\bauthor{\bsnm{Xu}, \binits{S.}},
\bauthor{\bsnm{Vaitla}, \binits{B.}},
\bauthor{\bsnm{Gonz{\'a}lez}, \binits{M.C.}}:
\batitle{Sequences of purchases in credit card data reveal lifestyles in urban
  populations}.
\bjtitle{Nature communications}
\bvolume{9}(\bissue{1}),
\bfpage{1}--\blpage{8}
(\byear{2018})
\end{barticle}
\endbibitem

\bibitem{toole2015path}
\begin{barticle}
\bauthor{\bsnm{Toole}, \binits{J.L.}},
\bauthor{\bsnm{Colak}, \binits{S.}},
\bauthor{\bsnm{Sturt}, \binits{B.}},
\bauthor{\bsnm{Alexander}, \binits{L.P.}},
\bauthor{\bsnm{Evsukoff}, \binits{A.}},
\bauthor{\bsnm{Gonz{\'a}lez}, \binits{M.C.}}:
\batitle{The path most traveled: Travel demand estimation using big data
  resources}.
\bjtitle{Transportation Research Part C: Emerging Technologies}
\bvolume{58},
\bfpage{162}--\blpage{177}
(\byear{2015})
\end{barticle}
\endbibitem

\bibitem{schneider2013unravelling}
\begin{barticle}
\bauthor{\bsnm{Schneider}, \binits{C.M.}},
\bauthor{\bsnm{Belik}, \binits{V.}},
\bauthor{\bsnm{Couronn{\'e}}, \binits{T.}},
\bauthor{\bsnm{Smoreda}, \binits{Z.}},
\bauthor{\bsnm{Gonz{\'a}lez}, \binits{M.C.}}:
\batitle{Unravelling daily human mobility motifs}.
\bjtitle{Journal of The Royal Society Interface}
\bvolume{10}(\bissue{84}),
\bfpage{20130246}
(\byear{2013})
\end{barticle}
\endbibitem

\bibitem{xu2018planning}
\begin{barticle}
\bauthor{\bsnm{Xu}, \binits{Y.}},
\bauthor{\bsnm{{\c{C}}olak}, \binits{S.}},
\bauthor{\bsnm{Kara}, \binits{E.C.}},
\bauthor{\bsnm{Moura}, \binits{S.J.}},
\bauthor{\bsnm{Gonz{\'a}lez}, \binits{M.C.}}:
\batitle{Planning for electric vehicle needs by coupling charging profiles with
  urban mobility}.
\bjtitle{Nature Energy}
\bvolume{3},
\bfpage{484}--\blpage{493}
(\byear{2018})
\end{barticle}
\endbibitem

\bibitem{rashidi2017exploring}
\begin{barticle}
\bauthor{\bsnm{Rashidi}, \binits{T.H.}},
\bauthor{\bsnm{Abbasi}, \binits{A.}},
\bauthor{\bsnm{Maghrebi}, \binits{M.}},
\bauthor{\bsnm{Hasan}, \binits{S.}},
\bauthor{\bsnm{Waller}, \binits{T.S.}}:
\batitle{Exploring the capacity of social media data for modelling travel
  behaviour: Opportunities and challenges}.
\bjtitle{Transportation Research Part C: Emerging Technologies}
\bvolume{75},
\bfpage{197}--\blpage{211}
(\byear{2017})
\end{barticle}
\endbibitem

\bibitem{scherrer2018travelers}
\begin{barticle}
\bauthor{\bsnm{Scherrer}, \binits{L.}},
\bauthor{\bsnm{Tomko}, \binits{M.}},
\bauthor{\bsnm{Ranacher}, \binits{P.}},
\bauthor{\bsnm{Weibel}, \binits{R.}}:
\batitle{{Travelers or locals? Identifying meaningful sub-populations from
  human movement data in the absence of ground truth}}.
\bjtitle{EPJ Data Science}
\bvolume{7}(\bissue{1}),
\bfpage{19}
(\byear{2018})
\end{barticle}
\endbibitem

\bibitem{dong2019segregated}
\begin{botherref}
\oauthor{\bsnm{Dong}, \binits{X.}},
\oauthor{\bsnm{Morales}, \binits{A.J.}},
\oauthor{\bsnm{Jahani}, \binits{E.}},
\oauthor{\bsnm{Moro}, \binits{E.}},
\oauthor{\bsnm{Lepri}, \binits{B.}},
\oauthor{\bsnm{Bozkaya}, \binits{B.}},
\oauthor{\bsnm{Sarraute}, \binits{C.}},
\oauthor{\bsnm{Bar-Yam}, \binits{Y.}},
\oauthor{\bsnm{Pentland}, \binits{A.}}:
Segregated interactions in urban and online spaces.
arXiv preprint arXiv:1911.04027
(2019)
\end{botherref}
\endbibitem

\bibitem{liao2019individual}
\begin{barticle}
\bauthor{\bsnm{Liao}, \binits{Y.}},
\bauthor{\bsnm{Yeh}, \binits{S.}},
\bauthor{\bsnm{Jeuken}, \binits{G.S.}}:
\batitle{From individual to collective behaviours: exploring population
  heterogeneity of human mobility based on social media data}.
\bjtitle{EPJ Data Science}
\bvolume{8}(\bissue{1}),
\bfpage{34}
(\byear{2019})
\end{barticle}
\endbibitem

\bibitem{mcneill2017estimating}
\begin{barticle}
\bauthor{\bsnm{McNeill}, \binits{G.}},
\bauthor{\bsnm{Bright}, \binits{J.}},
\bauthor{\bsnm{Hale}, \binits{S.A.}}:
\batitle{Estimating local commuting patterns from geolocated twitter data}.
\bjtitle{EPJ Data Science}
\bvolume{6}(\bissue{1}),
\bfpage{24}
(\byear{2017})
\end{barticle}
\endbibitem

\bibitem{aleta2020modeling}
\begin{botherref}
\oauthor{\bsnm{Aleta}, \binits{A.}},
\oauthor{\bparticle{y} \bsnm{Piontti}, \binits{A.P.}},
\oauthor{\bsnm{Ajelli}, \binits{M.}},
\oauthor{\bsnm{Litvinova}, \binits{M.}}, et al.:
Modeling the impact of social distancing, testing, contact tracing and
  household quarantine on second-wave scen-arios of the covid-19 epidemic.
Technical report
\end{botherref}
\endbibitem

\bibitem{klein2020assessing}
\begin{botherref}
\oauthor{\bsnm{Klein}, \binits{B.}},
\oauthor{\bsnm{Privitera}, \binits{F.}},
\oauthor{\bsnm{Lake}, \binits{B.}},
\oauthor{\bsnm{Kraemer}, \binits{M.U.}},
\oauthor{\bsnm{Brownstein}, \binits{J.S.}},
\oauthor{\bsnm{Lazer}, \binits{D.}},
\oauthor{\bsnm{Eliassi-Rad}, \binits{T.}}, et al.:
Assessing changes in commuting and individual mobility in major metropolitan
  areas in the United States during the COVID-19 outbreak
(2020)
\end{botherref}
\endbibitem

\bibitem{kwan2004gis}
\begin{barticle}
\bauthor{\bsnm{Kwan}, \binits{M.-P.}}:
\batitle{{GIS methods in time-geographic research: Geocomputation and
  geovisualization of human activity patterns}}.
\bjtitle{Geografiska Annaler: Series B, Human Geography}
\bvolume{86}(\bissue{4}),
\bfpage{267}--\blpage{280}
(\byear{2004})
\end{barticle}
\endbibitem

\bibitem{ratti2006mobile}
\begin{barticle}
\bauthor{\bsnm{Ratti}, \binits{C.}},
\bauthor{\bsnm{Frenchman}, \binits{D.}},
\bauthor{\bsnm{Pulselli}, \binits{R.M.}},
\bauthor{\bsnm{Williams}, \binits{S.}}:
\batitle{Mobile landscapes: Using location data from cell phones for urban
  analysis}.
\bjtitle{Environment and Planning B: Planning and Design}
\bvolume{33}(\bissue{5}),
\bfpage{727}--\blpage{748}
(\byear{2006})
\end{barticle}
\endbibitem

\bibitem{zheng2015trajectory}
\begin{barticle}
\bauthor{\bsnm{Zheng}, \binits{Y.}}:
\batitle{Trajectory data mining: An overview}.
\bjtitle{ACM Transactions on Intelligent Systems and Technology (TIST)}
\bvolume{6}(\bissue{3}),
\bfpage{29}
(\byear{2015})
\end{barticle}
\endbibitem

\bibitem{miller2015data}
\begin{barticle}
\bauthor{\bsnm{Miller}, \binits{H.J.}},
\bauthor{\bsnm{Goodchild}, \binits{M.F.}}:
\batitle{Data-driven geography}.
\bjtitle{GeoJournal}
\bvolume{80}(\bissue{4}),
\bfpage{449}--\blpage{461}
(\byear{2015})
\end{barticle}
\endbibitem

\bibitem{silva2019urban}
\begin{barticle}
\bauthor{\bsnm{Silva}, \binits{T.H.}},
\bauthor{\bsnm{Viana}, \binits{A.C.}},
\bauthor{\bsnm{Benevenuto}, \binits{F.}},
\bauthor{\bsnm{Villas}, \binits{L.}},
\bauthor{\bsnm{Salles}, \binits{J.}},
\bauthor{\bsnm{Loureiro}, \binits{A.}},
\bauthor{\bsnm{Quercia}, \binits{D.}}:
\batitle{Urban computing leveraging location-based social network data: A
  survey}.
\bjtitle{ACM Computing Surveys (CSUR)}
\bvolume{52}(\bissue{1}),
\bfpage{17}
(\byear{2019})
\end{barticle}
\endbibitem

\bibitem{cuebiq2019}
\begin{botherref}
\oauthor{\bsnm{{Cuebiq Offline Intelligence Measurement}}}
\url{https://www.cuebiq.com}.
[Online; accessed September-2019]
(2019)
\end{botherref}
\endbibitem

\bibitem{xiao2015travel}
\begin{barticle}
\bauthor{\bsnm{Xiao}, \binits{G.}},
\bauthor{\bsnm{Juan}, \binits{Z.}},
\bauthor{\bsnm{Zhang}, \binits{C.}}:
\batitle{{Travel mode detection based on GPS track data and Bayesian
  networks}}.
\bjtitle{Computers, Environment and Urban Systems}
\bvolume{54},
\bfpage{14}--\blpage{22}
(\byear{2015})
\end{barticle}
\endbibitem

\bibitem{dabiri2018inferring}
\begin{barticle}
\bauthor{\bsnm{Dabiri}, \binits{S.}},
\bauthor{\bsnm{Heaslip}, \binits{K.}}:
\batitle{{Inferring transportation modes from GPS trajectories using a
  convolutional neural network}}.
\bjtitle{Transportation Research Part C: Emerging Technologies}
\bvolume{86},
\bfpage{360}--\blpage{371}
(\byear{2018})
\end{barticle}
\endbibitem

\bibitem{jiang2013review}
\begin{bchapter}
\bauthor{\bsnm{Jiang}, \binits{S.}},
\bauthor{\bsnm{Fiore}, \binits{G.A.}},
\bauthor{\bsnm{Yang}, \binits{Y.}},
\bauthor{\bsnm{Ferreira~Jr}, \binits{J.}},
\bauthor{\bsnm{Frazzoli}, \binits{E.}},
\bauthor{\bsnm{Gonz{\'a}lez}, \binits{M.C.}}:
\bctitle{A review of urban computing for mobile phone traces: Current methods,
  challenges and opportunities}.
In: \bbtitle{Proceedings of the 2nd ACM SIGKDD International Workshop on Urban
  Computing},
p. \bfpage{2}
(\byear{2013}).
\bcomment{ACM}
\end{bchapter}
\endbibitem

\bibitem{ccolak2015analyzing}
\begin{botherref}
\oauthor{\bsnm{{\c{C}}olak}, \binits{S.}},
\oauthor{\bsnm{Alexander}, \binits{L.P.}},
\oauthor{\bsnm{Alvim}, \binits{B.G.}},
\oauthor{\bsnm{Mehndiratta}, \binits{S.R.}},
\oauthor{\bsnm{Gonz{\'a}lez}, \binits{M.C.}}:
Analyzing cell phone location data for urban travel: current methods,
  limitations, and opportunities.
Transportation research record: Journal of the transportation research board
(2526),
126--135
(2015)
\end{botherref}
\endbibitem

\bibitem{vanhoof2018assessing}
\begin{barticle}
\bauthor{\bsnm{Vanhoof}, \binits{M.}},
\bauthor{\bsnm{Reis}, \binits{F.}},
\bauthor{\bsnm{Ploetz}, \binits{T.}},
\bauthor{\bsnm{Smoreda}, \binits{Z.}}:
\batitle{Assessing the quality of home detection from mobile phone data for
  official statistics}.
\bjtitle{Journal of Official Statistics}
\bvolume{34}(\bissue{4}),
\bfpage{935}--\blpage{960}
(\byear{2018})
\end{barticle}
\endbibitem

\bibitem{census2016}
\begin{botherref}
\oauthor{\bsnm{{U.S. Census Bureau}}}
\url{https://www.census.gov/}.
[Online; accessed September-2018]
(2016)
\end{botherref}
\endbibitem

\bibitem{nctcog2014}
\begin{botherref}
\oauthor{\bsnm{{The North Central Texas Council of Governments}}}
\url{https://www.nctcog.org/}.
[Online; accessed September-2018]
(2014)
\end{botherref}
\endbibitem

\bibitem{sun2015bayesian}
\begin{barticle}
\bauthor{\bsnm{Sun}, \binits{L.}},
\bauthor{\bsnm{Erath}, \binits{A.}}:
\batitle{A bayesian network approach for population synthesis}.
\bjtitle{Transportation Research Part C: Emerging Technologies}
\bvolume{61},
\bfpage{49}--\blpage{62}
(\byear{2015})
\end{barticle}
\endbibitem

\bibitem{chen2014map}
\begin{barticle}
\bauthor{\bsnm{Chen}, \binits{B.Y.}},
\bauthor{\bsnm{Yuan}, \binits{H.}},
\bauthor{\bsnm{Li}, \binits{Q.}},
\bauthor{\bsnm{Lam}, \binits{W.H.}},
\bauthor{\bsnm{Shaw}, \binits{S.-L.}},
\bauthor{\bsnm{Yan}, \binits{K.}}:
\batitle{Map-matching algorithm for large-scale low-frequency floating car
  data}.
\bjtitle{International Journal of Geographical Information Science}
\bvolume{28}(\bissue{1}),
\bfpage{22}--\blpage{38}
(\byear{2014})
\end{barticle}
\endbibitem

\bibitem{kim2015spatial}
\begin{barticle}
\bauthor{\bsnm{Kim}, \binits{J.}},
\bauthor{\bsnm{Mahmassani}, \binits{H.S.}}:
\batitle{Spatial and temporal characterization of travel patterns in a traffic
  network using vehicle trajectories}.
\bjtitle{Transportation Research Part C: Emerging Technologies}
\bvolume{59},
\bfpage{375}--\blpage{390}
(\byear{2015})
\end{barticle}
\endbibitem

\bibitem{atev2010clustering}
\begin{barticle}
\bauthor{\bsnm{Atev}, \binits{S.}},
\bauthor{\bsnm{Miller}, \binits{G.}},
\bauthor{\bsnm{Papanikolopoulos}, \binits{N.P.}}:
\batitle{Clustering of vehicle trajectories}.
\bjtitle{IEEE Transactions on Intelligent Transportation Systems}
\bvolume{11}(\bissue{3}),
\bfpage{647}--\blpage{657}
(\byear{2010})
\end{barticle}
\endbibitem

\bibitem{TTRreport}
\begin{botherref}
\oauthor{\bsnm{{FHWA}}}
\url{https://ops.fhwa.dot.gov/publications/tt_reliability/TTR_Report.htm}.
[Online; accessed September-2019]
(2019)
\end{botherref}
\endbibitem

\end{thebibliography}

\newcommand{\BMCxmlcomment}[1]{}

\BMCxmlcomment{

<refgrp>

<bibl id="B1">
  <title><p>Deconstructing laws of accessibility and facility distribution in
  cities</p></title>
  <aug>
    <au><snm>Xu</snm><fnm>Y</fnm></au>
    <au><snm>Olmos</snm><fnm>LE</fnm></au>
    <au><snm>Abbar</snm><fnm>S</fnm></au>
    <au><snm>Gonz{\'a}lez</snm><fnm>MC</fnm></au>
  </aug>
  <source>Science advances</source>
  <publisher>American Association for the Advancement of Science</publisher>
  <pubdate>2020</pubdate>
  <volume>6</volume>
  <issue>37</issue>
  <fpage>eabb4112</fpage>
</bibl>

<bibl id="B2">
  <title><p>Measuring economic costs of urban traffic congestion to
  business</p></title>
  <aug>
    <au><snm>Weisbrod</snm><fnm>G</fnm></au>
    <au><snm>Vary</snm><fnm>D</fnm></au>
    <au><snm>Treyz</snm><fnm>G</fnm></au>
  </aug>
  <source>Transportation Research Record: Journal of the Transportation
  Research Board</source>
  <publisher>Transportation Research Board of the National
  Academies</publisher>
  <pubdate>2003</pubdate>
  <issue>1839</issue>
  <fpage>98</fpage>
  <lpage>-106</lpage>
</bibl>

<bibl id="B3">
  <title><p>{Transport and public health in China: the road to a healthy
  future}</p></title>
  <aug>
    <au><snm>Jiang</snm><fnm>B</fnm></au>
    <au><snm>Liang</snm><fnm>S</fnm></au>
    <au><snm>Peng</snm><fnm>ZR</fnm></au>
    <au><snm>Cong</snm><fnm>H</fnm></au>
    <au><snm>Levy</snm><fnm>M</fnm></au>
    <au><snm>Cheng</snm><fnm>Q</fnm></au>
    <au><snm>Wang</snm><fnm>T</fnm></au>
    <au><snm>Remais</snm><fnm>JV</fnm></au>
  </aug>
  <source>The Lancet</source>
  <publisher>Elsevier</publisher>
  <pubdate>2017</pubdate>
  <volume>390</volume>
  <issue>10104</issue>
  <fpage>1781</fpage>
  <lpage>-1791</lpage>
</bibl>

<bibl id="B4">
  <title><p>{Unraveling environmental justice in ambient PM$_{2.5}$ exposure in
  Beijing: A big data approach}</p></title>
  <aug>
    <au><snm>Xu</snm><fnm>Y</fnm></au>
    <au><snm>Jiang</snm><fnm>S</fnm></au>
    <au><snm>Li</snm><fnm>R</fnm></au>
    <au><snm>Zhang</snm><fnm>J</fnm></au>
    <au><snm>Zhao</snm><fnm>J</fnm></au>
    <au><snm>Abbar</snm><fnm>S</fnm></au>
    <au><snm>Gonz{\'a}lez</snm><fnm>MC</fnm></au>
  </aug>
  <source>Computers, Environment and Urban Systems</source>
  <publisher>Elsevier</publisher>
  <pubdate>2019</pubdate>
  <volume>75</volume>
  <fpage>12</fpage>
  <lpage>-21</lpage>
</bibl>

<bibl id="B5">
  <title><p>Day-to-day evolution of network flows under real-time information
  and reactive signal control</p></title>
  <aug>
    <au><snm>Hu</snm><fnm>TY</fnm></au>
    <au><snm>Mahmassani</snm><fnm>HS</fnm></au>
  </aug>
  <source>Transportation Research Part C: Emerging Technologies</source>
  <publisher>Elsevier</publisher>
  <pubdate>1997</pubdate>
  <volume>5</volume>
  <issue>1</issue>
  <fpage>51</fpage>
  <lpage>-69</lpage>
</bibl>

<bibl id="B6">
  <title><p>Transport solutions for cleaner air</p></title>
  <aug>
    <au><snm>Kelly</snm><fnm>FJ</fnm></au>
    <au><snm>Zhu</snm><fnm>T</fnm></au>
  </aug>
  <source>Science</source>
  <publisher>American Association for the Advancement of Science</publisher>
  <pubdate>2016</pubdate>
  <volume>352</volume>
  <issue>6288</issue>
  <fpage>934</fpage>
  <lpage>-936</lpage>
</bibl>

<bibl id="B7">
  <title><p>Collective benefits in traffic during mega events via the use of
  information technologies</p></title>
  <aug>
    <au><snm>Xu</snm><fnm>Y</fnm></au>
    <au><snm>Gonz{\'a}lez</snm><fnm>MC</fnm></au>
  </aug>
  <source>Journal of The Royal Society Interface</source>
  <publisher>The Royal Society</publisher>
  <pubdate>2017</pubdate>
  <volume>14</volume>
  <issue>129</issue>
  <fpage>20161041</fpage>
</bibl>

<bibl id="B8">
  <title><p>Understanding congested travel in urban areas</p></title>
  <aug>
    <au><snm>{\c{C}}olak</snm><fnm>S</fnm></au>
    <au><snm>Lima</snm><fnm>A</fnm></au>
    <au><snm>Gonz{\'a}lez</snm><fnm>MC</fnm></au>
  </aug>
  <source>Nature communications</source>
  <publisher>Nature Publishing Group</publisher>
  <pubdate>2016</pubdate>
  <volume>7</volume>
  <issue>1</issue>
  <fpage>1</fpage>
  <lpage>-8</lpage>
</bibl>

<bibl id="B9">
  <title><p>The London congestion charge: a tentative economic
  appraisal</p></title>
  <aug>
    <au><snm>Prud'homme</snm><fnm>R</fnm></au>
    <au><snm>Bocarejo</snm><fnm>JP</fnm></au>
  </aug>
  <source>Transport Policy</source>
  <publisher>Elsevier</publisher>
  <pubdate>2005</pubdate>
  <volume>12</volume>
  <issue>3</issue>
  <fpage>279</fpage>
  <lpage>-287</lpage>
</bibl>

<bibl id="B10">
  <title><p>Stabilizing traffic with autonomous vehicles</p></title>
  <aug>
    <au><snm>Wu</snm><fnm>C</fnm></au>
    <au><snm>Bayen</snm><fnm>AM</fnm></au>
    <au><snm>Mehta</snm><fnm>A</fnm></au>
  </aug>
  <source>2018 IEEE International Conference on Robotics and Automation
  (ICRA)</source>
  <pubdate>2018</pubdate>
  <fpage>1</fpage>
  <lpage>-7</lpage>
</bibl>

<bibl id="B11">
  <title><p>Understanding individual human mobility patterns</p></title>
  <aug>
    <au><snm>Gonzalez</snm><fnm>MC</fnm></au>
    <au><snm>Hidalgo</snm><fnm>CA</fnm></au>
    <au><snm>Barabasi</snm><fnm>AL</fnm></au>
  </aug>
  <source>Nature</source>
  <publisher>Nature publishing group</publisher>
  <pubdate>2008</pubdate>
  <volume>453</volume>
  <issue>7196</issue>
  <fpage>779</fpage>
</bibl>

<bibl id="B12">
  <title><p>Returners and explorers dichotomy in human mobility</p></title>
  <aug>
    <au><snm>Pappalardo</snm><fnm>L</fnm></au>
    <au><snm>Simini</snm><fnm>F</fnm></au>
    <au><snm>Rinzivillo</snm><fnm>S</fnm></au>
    <au><snm>Pedreschi</snm><fnm>D</fnm></au>
    <au><snm>Giannotti</snm><fnm>F</fnm></au>
    <au><snm>Barab{\'a}si</snm><fnm>AL</fnm></au>
  </aug>
  <source>Nature communications</source>
  <publisher>Nature Publishing Group</publisher>
  <pubdate>2015</pubdate>
  <volume>6</volume>
  <fpage>8166</fpage>
</bibl>

<bibl id="B13">
  <title><p>{The TimeGeo modeling framework for urban mobility without travel
  surveys}</p></title>
  <aug>
    <au><snm>Jiang</snm><fnm>S</fnm></au>
    <au><snm>Yang</snm><fnm>Y</fnm></au>
    <au><snm>Gupta</snm><fnm>S</fnm></au>
    <au><snm>Veneziano</snm><fnm>D</fnm></au>
    <au><snm>Athavale</snm><fnm>S</fnm></au>
    <au><snm>Gonz{\'a}lez</snm><fnm>MC</fnm></au>
  </aug>
  <source>Proceedings of the National Academy of Sciences</source>
  <publisher>National Acad Sciences</publisher>
  <pubdate>2016</pubdate>
  <volume>113</volume>
  <issue>37</issue>
  <fpage>E5370</fpage>
  <lpage>-E5378</lpage>
</bibl>

<bibl id="B14">
  <title><p>Discrete choice methods and their applications to short term travel
  decisions</p></title>
  <aug>
    <au><snm>Ben Akiva</snm><fnm>M</fnm></au>
    <au><snm>Bierlaire</snm><fnm>M</fnm></au>
  </aug>
  <source>Handbook of Transportation Science</source>
  <publisher>Boston, MA: Springer</publisher>
  <pubdate>1999</pubdate>
  <fpage>5</fpage>
  <lpage>-33</lpage>
</bibl>

<bibl id="B15">
  <title><p>Route choice modeling: past, present and future research
  directions</p></title>
  <aug>
    <au><snm>Prato</snm><fnm>CG</fnm></au>
  </aug>
  <source>Journal of choice modelling</source>
  <publisher>Elsevier</publisher>
  <pubdate>2009</pubdate>
  <volume>2</volume>
  <issue>1</issue>
  <fpage>65</fpage>
  <lpage>-100</lpage>
</bibl>

<bibl id="B16">
  <title><p>Boundedly rational route choice behavior: A review of models and
  methodologies</p></title>
  <aug>
    <au><snm>Di</snm><fnm>X</fnm></au>
    <au><snm>Liu</snm><fnm>HX</fnm></au>
  </aug>
  <source>Transportation Research Part B: Methodological</source>
  <publisher>Elsevier</publisher>
  <pubdate>2016</pubdate>
  <volume>85</volume>
  <fpage>142</fpage>
  <lpage>-179</lpage>
</bibl>

<bibl id="B17">
  <title><p>Do people use the shortest path? An empirical test of Wardrop’s
  first principle</p></title>
  <aug>
    <au><snm>Zhu</snm><fnm>S</fnm></au>
    <au><snm>Levinson</snm><fnm>D</fnm></au>
  </aug>
  <source>PloS one</source>
  <publisher>Public Library of Science</publisher>
  <pubdate>2015</pubdate>
  <volume>10</volume>
  <issue>8</issue>
  <fpage>e0134322</fpage>
</bibl>

<bibl id="B18">
  <title><p>Understanding individual routing behaviour</p></title>
  <aug>
    <au><snm>Lima</snm><fnm>A</fnm></au>
    <au><snm>Stanojevic</snm><fnm>R</fnm></au>
    <au><snm>Papagiannaki</snm><fnm>D</fnm></au>
    <au><snm>Rodriguez</snm><fnm>P</fnm></au>
    <au><snm>Gonz{\'a}lez</snm><fnm>MC</fnm></au>
  </aug>
  <source>Journal of The Royal Society Interface</source>
  <publisher>The Royal Society</publisher>
  <pubdate>2016</pubdate>
  <volume>13</volume>
  <issue>116</issue>
  <fpage>20160021</fpage>
</bibl>

<bibl id="B19">
  <title><p>Sequences of purchases in credit card data reveal lifestyles in
  urban populations</p></title>
  <aug>
    <au><snm>Di Clemente</snm><fnm>R</fnm></au>
    <au><snm>Luengo Oroz</snm><fnm>M</fnm></au>
    <au><snm>Travizano</snm><fnm>M</fnm></au>
    <au><snm>Xu</snm><fnm>S</fnm></au>
    <au><snm>Vaitla</snm><fnm>B</fnm></au>
    <au><snm>Gonz{\'a}lez</snm><fnm>MC</fnm></au>
  </aug>
  <source>Nature communications</source>
  <publisher>Nature Publishing Group</publisher>
  <pubdate>2018</pubdate>
  <volume>9</volume>
  <issue>1</issue>
  <fpage>1</fpage>
  <lpage>-8</lpage>
</bibl>

<bibl id="B20">
  <title><p>The path most traveled: Travel demand estimation using big data
  resources</p></title>
  <aug>
    <au><snm>Toole</snm><fnm>JL</fnm></au>
    <au><snm>Colak</snm><fnm>S</fnm></au>
    <au><snm>Sturt</snm><fnm>B</fnm></au>
    <au><snm>Alexander</snm><fnm>LP</fnm></au>
    <au><snm>Evsukoff</snm><fnm>A</fnm></au>
    <au><snm>Gonz{\'a}lez</snm><fnm>MC</fnm></au>
  </aug>
  <source>Transportation Research Part C: Emerging Technologies</source>
  <publisher>Elsevier</publisher>
  <pubdate>2015</pubdate>
  <volume>58</volume>
  <fpage>162</fpage>
  <lpage>-177</lpage>
</bibl>

<bibl id="B21">
  <title><p>Unravelling daily human mobility motifs</p></title>
  <aug>
    <au><snm>Schneider</snm><fnm>CM</fnm></au>
    <au><snm>Belik</snm><fnm>V</fnm></au>
    <au><snm>Couronn{\'e}</snm><fnm>T</fnm></au>
    <au><snm>Smoreda</snm><fnm>Z</fnm></au>
    <au><snm>Gonz{\'a}lez</snm><fnm>MC</fnm></au>
  </aug>
  <source>Journal of The Royal Society Interface</source>
  <publisher>The Royal Society</publisher>
  <pubdate>2013</pubdate>
  <volume>10</volume>
  <issue>84</issue>
  <fpage>20130246</fpage>
</bibl>

<bibl id="B22">
  <title><p>Planning for electric vehicle needs by coupling charging profiles
  with urban mobility</p></title>
  <aug>
    <au><snm>Xu</snm><fnm>Y</fnm></au>
    <au><snm>{\c{C}}olak</snm><fnm>S</fnm></au>
    <au><snm>Kara</snm><fnm>EC</fnm></au>
    <au><snm>Moura</snm><fnm>SJ</fnm></au>
    <au><snm>Gonz{\'a}lez</snm><fnm>MC</fnm></au>
  </aug>
  <source>Nature Energy</source>
  <publisher>Nature Publishing Group</publisher>
  <pubdate>2018</pubdate>
  <volume>3</volume>
  <fpage>484</fpage>
  <lpage>-493</lpage>
</bibl>

<bibl id="B23">
  <title><p>Exploring the capacity of social media data for modelling travel
  behaviour: Opportunities and challenges</p></title>
  <aug>
    <au><snm>Rashidi</snm><fnm>TH</fnm></au>
    <au><snm>Abbasi</snm><fnm>A</fnm></au>
    <au><snm>Maghrebi</snm><fnm>M</fnm></au>
    <au><snm>Hasan</snm><fnm>S</fnm></au>
    <au><snm>Waller</snm><fnm>TS</fnm></au>
  </aug>
  <source>Transportation Research Part C: Emerging Technologies</source>
  <publisher>Elsevier</publisher>
  <pubdate>2017</pubdate>
  <volume>75</volume>
  <fpage>197</fpage>
  <lpage>-211</lpage>
</bibl>

<bibl id="B24">
  <title><p>{Travelers or locals? Identifying meaningful sub-populations from
  human movement data in the absence of ground truth}</p></title>
  <aug>
    <au><snm>Scherrer</snm><fnm>L</fnm></au>
    <au><snm>Tomko</snm><fnm>M</fnm></au>
    <au><snm>Ranacher</snm><fnm>P</fnm></au>
    <au><snm>Weibel</snm><fnm>R</fnm></au>
  </aug>
  <source>EPJ Data Science</source>
  <publisher>SpringerOpen</publisher>
  <pubdate>2018</pubdate>
  <volume>7</volume>
  <issue>1</issue>
  <fpage>19</fpage>
</bibl>

<bibl id="B25">
  <title><p>Segregated interactions in urban and online spaces</p></title>
  <aug>
    <au><snm>Dong</snm><fnm>X</fnm></au>
    <au><snm>Morales</snm><fnm>AJ</fnm></au>
    <au><snm>Jahani</snm><fnm>E</fnm></au>
    <au><snm>Moro</snm><fnm>E</fnm></au>
    <au><snm>Lepri</snm><fnm>B</fnm></au>
    <au><snm>Bozkaya</snm><fnm>B</fnm></au>
    <au><snm>Sarraute</snm><fnm>C</fnm></au>
    <au><snm>Bar Yam</snm><fnm>Y</fnm></au>
    <au><snm>Pentland</snm><fnm>A</fnm></au>
  </aug>
  <source>arXiv preprint arXiv:1911.04027</source>
  <pubdate>2019</pubdate>
</bibl>

<bibl id="B26">
  <title><p>From individual to collective behaviours: exploring population
  heterogeneity of human mobility based on social media data</p></title>
  <aug>
    <au><snm>Liao</snm><fnm>Y</fnm></au>
    <au><snm>Yeh</snm><fnm>S</fnm></au>
    <au><snm>Jeuken</snm><fnm>GS</fnm></au>
  </aug>
  <source>EPJ Data Science</source>
  <publisher>Springer Berlin Heidelberg</publisher>
  <pubdate>2019</pubdate>
  <volume>8</volume>
  <issue>1</issue>
  <fpage>34</fpage>
</bibl>

<bibl id="B27">
  <title><p>Estimating local commuting patterns from geolocated Twitter
  data</p></title>
  <aug>
    <au><snm>McNeill</snm><fnm>G</fnm></au>
    <au><snm>Bright</snm><fnm>J</fnm></au>
    <au><snm>Hale</snm><fnm>SA</fnm></au>
  </aug>
  <source>EPJ Data Science</source>
  <publisher>Springer</publisher>
  <pubdate>2017</pubdate>
  <volume>6</volume>
  <issue>1</issue>
  <fpage>24</fpage>
</bibl>

<bibl id="B28">
  <title><p>Modeling the impact of social distancing, testing, contact tracing
  and household quarantine on second-wave scen-arios of the COVID-19
  epidemic</p></title>
  <aug>
    <au><snm>Aleta</snm><fnm>A</fnm></au>
    <au><snm>Piontti</snm><fnm>AP</fnm></au>
    <au><snm>Ajelli</snm><fnm>M</fnm></au>
    <au><snm>Litvinova</snm><fnm>M</fnm></au>
    <au><cnm>others</cnm></au>
  </aug>
</bibl>

<bibl id="B29">
  <title><p>Assessing changes in commuting and individual mobility in major
  metropolitan areas in the United States during the COVID-19
  outbreak</p></title>
  <aug>
    <au><snm>Klein</snm><fnm>B</fnm></au>
    <au><snm>Privitera</snm><fnm>F</fnm></au>
    <au><snm>Lake</snm><fnm>B</fnm></au>
    <au><snm>Kraemer</snm><fnm>MU</fnm></au>
    <au><snm>Brownstein</snm><fnm>JS</fnm></au>
    <au><snm>Lazer</snm><fnm>D</fnm></au>
    <au><snm>Eliassi Rad</snm><fnm>T</fnm></au>
    <au><cnm>others</cnm></au>
  </aug>
  <pubdate>2020</pubdate>
</bibl>

<bibl id="B30">
  <title><p>{GIS methods in time-geographic research: Geocomputation and
  geovisualization of human activity patterns}</p></title>
  <aug>
    <au><snm>Kwan</snm><fnm>MP</fnm></au>
  </aug>
  <source>Geografiska Annaler: Series B, Human Geography</source>
  <publisher>Wiley Online Library</publisher>
  <pubdate>2004</pubdate>
  <volume>86</volume>
  <issue>4</issue>
  <fpage>267</fpage>
  <lpage>-280</lpage>
</bibl>

<bibl id="B31">
  <title><p>Mobile landscapes: Using location data from cell phones for urban
  analysis</p></title>
  <aug>
    <au><snm>Ratti</snm><fnm>C</fnm></au>
    <au><snm>Frenchman</snm><fnm>D</fnm></au>
    <au><snm>Pulselli</snm><fnm>RM</fnm></au>
    <au><snm>Williams</snm><fnm>S</fnm></au>
  </aug>
  <source>Environment and Planning B: Planning and Design</source>
  <publisher>SAGE Publications Sage UK: London, England</publisher>
  <pubdate>2006</pubdate>
  <volume>33</volume>
  <issue>5</issue>
  <fpage>727</fpage>
  <lpage>-748</lpage>
</bibl>

<bibl id="B32">
  <title><p>Trajectory data mining: An overview</p></title>
  <aug>
    <au><snm>Zheng</snm><fnm>Y</fnm></au>
  </aug>
  <source>ACM Transactions on Intelligent Systems and Technology
  (TIST)</source>
  <publisher>ACM</publisher>
  <pubdate>2015</pubdate>
  <volume>6</volume>
  <issue>3</issue>
  <fpage>29</fpage>
</bibl>

<bibl id="B33">
  <title><p>Data-driven geography</p></title>
  <aug>
    <au><snm>Miller</snm><fnm>HJ</fnm></au>
    <au><snm>Goodchild</snm><fnm>MF</fnm></au>
  </aug>
  <source>GeoJournal</source>
  <publisher>Springer</publisher>
  <pubdate>2015</pubdate>
  <volume>80</volume>
  <issue>4</issue>
  <fpage>449</fpage>
  <lpage>-461</lpage>
</bibl>

<bibl id="B34">
  <title><p>Urban Computing Leveraging Location-Based Social Network Data: A
  Survey</p></title>
  <aug>
    <au><snm>Silva</snm><fnm>TH</fnm></au>
    <au><snm>Viana</snm><fnm>AC</fnm></au>
    <au><snm>Benevenuto</snm><fnm>F</fnm></au>
    <au><snm>Villas</snm><fnm>L</fnm></au>
    <au><snm>Salles</snm><fnm>J</fnm></au>
    <au><snm>Loureiro</snm><fnm>A</fnm></au>
    <au><snm>Quercia</snm><fnm>D</fnm></au>
  </aug>
  <source>ACM Computing Surveys (CSUR)</source>
  <publisher>ACM</publisher>
  <pubdate>2019</pubdate>
  <volume>52</volume>
  <issue>1</issue>
  <fpage>17</fpage>
</bibl>

<bibl id="B35">
  <aug>
    <au><cnm>{Cuebiq Offline Intelligence Measurement}</cnm></au>
  </aug>
  <source>\url{https://www.cuebiq.com}</source>
  <pubdate>2019</pubdate>
  <note>[Online; accessed September-2019]</note>
</bibl>

<bibl id="B36">
  <title><p>{Travel mode detection based on GPS track data and Bayesian
  networks}</p></title>
  <aug>
    <au><snm>Xiao</snm><fnm>G</fnm></au>
    <au><snm>Juan</snm><fnm>Z</fnm></au>
    <au><snm>Zhang</snm><fnm>C</fnm></au>
  </aug>
  <source>Computers, Environment and Urban Systems</source>
  <publisher>Elsevier</publisher>
  <pubdate>2015</pubdate>
  <volume>54</volume>
  <fpage>14</fpage>
  <lpage>-22</lpage>
</bibl>

<bibl id="B37">
  <title><p>{Inferring transportation modes from GPS trajectories using a
  convolutional neural network}</p></title>
  <aug>
    <au><snm>Dabiri</snm><fnm>S</fnm></au>
    <au><snm>Heaslip</snm><fnm>K</fnm></au>
  </aug>
  <source>Transportation Research Part C: Emerging Technologies</source>
  <publisher>Elsevier</publisher>
  <pubdate>2018</pubdate>
  <volume>86</volume>
  <fpage>360</fpage>
  <lpage>-371</lpage>
</bibl>

<bibl id="B38">
  <title><p>A review of urban computing for mobile phone traces: Current
  methods, challenges and opportunities</p></title>
  <aug>
    <au><snm>Jiang</snm><fnm>S</fnm></au>
    <au><snm>Fiore</snm><fnm>GA</fnm></au>
    <au><snm>Yang</snm><fnm>Y</fnm></au>
    <au><snm>Ferreira Jr</snm><fnm>J</fnm></au>
    <au><snm>Frazzoli</snm><fnm>E</fnm></au>
    <au><snm>Gonz{\'a}lez</snm><fnm>MC</fnm></au>
  </aug>
  <source>Proceedings of the 2nd ACM SIGKDD International Workshop on Urban
  Computing</source>
  <pubdate>2013</pubdate>
  <fpage>2</fpage>
</bibl>

<bibl id="B39">
  <title><p>Analyzing cell phone location data for urban travel: current
  methods, limitations, and opportunities</p></title>
  <aug>
    <au><snm>{\c{C}}olak</snm><fnm>S</fnm></au>
    <au><snm>Alexander</snm><fnm>LP</fnm></au>
    <au><snm>Alvim</snm><fnm>BG</fnm></au>
    <au><snm>Mehndiratta</snm><fnm>SR</fnm></au>
    <au><snm>Gonz{\'a}lez</snm><fnm>MC</fnm></au>
  </aug>
  <source>Transportation research record: Journal of the transportation
  research board</source>
  <publisher>Transportation Research Board of the National
  Academies</publisher>
  <pubdate>2015</pubdate>
  <issue>2526</issue>
  <fpage>126</fpage>
  <lpage>-135</lpage>
</bibl>

<bibl id="B40">
  <title><p>Assessing the quality of home detection from mobile phone data for
  official statistics</p></title>
  <aug>
    <au><snm>Vanhoof</snm><fnm>M</fnm></au>
    <au><snm>Reis</snm><fnm>F</fnm></au>
    <au><snm>Ploetz</snm><fnm>T</fnm></au>
    <au><snm>Smoreda</snm><fnm>Z</fnm></au>
  </aug>
  <source>Journal of Official Statistics</source>
  <publisher>Sciendo</publisher>
  <pubdate>2018</pubdate>
  <volume>34</volume>
  <issue>4</issue>
  <fpage>935</fpage>
  <lpage>-960</lpage>
</bibl>

<bibl id="B41">
  <aug>
    <au><cnm>{U.S. Census Bureau}</cnm></au>
  </aug>
  <source>\url{https://www.census.gov/}</source>
  <pubdate>2016</pubdate>
  <note>[Online; accessed September-2018]</note>
</bibl>

<bibl id="B42">
  <aug>
    <au><cnm>{The North Central Texas Council of Governments}</cnm></au>
  </aug>
  <source>\url{https://www.nctcog.org/}</source>
  <pubdate>2014</pubdate>
  <note>[Online; accessed September-2018]</note>
</bibl>

<bibl id="B43">
  <title><p>A Bayesian network approach for population synthesis</p></title>
  <aug>
    <au><snm>Sun</snm><fnm>L</fnm></au>
    <au><snm>Erath</snm><fnm>A</fnm></au>
  </aug>
  <source>Transportation Research Part C: Emerging Technologies</source>
  <publisher>Elsevier</publisher>
  <pubdate>2015</pubdate>
  <volume>61</volume>
  <fpage>49</fpage>
  <lpage>-62</lpage>
</bibl>

<bibl id="B44">
  <title><p>Map-matching algorithm for large-scale low-frequency floating car
  data</p></title>
  <aug>
    <au><snm>Chen</snm><fnm>BY</fnm></au>
    <au><snm>Yuan</snm><fnm>H</fnm></au>
    <au><snm>Li</snm><fnm>Q</fnm></au>
    <au><snm>Lam</snm><fnm>WH</fnm></au>
    <au><snm>Shaw</snm><fnm>SL</fnm></au>
    <au><snm>Yan</snm><fnm>K</fnm></au>
  </aug>
  <source>International Journal of Geographical Information Science</source>
  <publisher>Taylor \& Francis</publisher>
  <pubdate>2014</pubdate>
  <volume>28</volume>
  <issue>1</issue>
  <fpage>22</fpage>
  <lpage>-38</lpage>
</bibl>

<bibl id="B45">
  <title><p>Spatial and temporal characterization of travel patterns in a
  traffic network using vehicle trajectories</p></title>
  <aug>
    <au><snm>Kim</snm><fnm>J</fnm></au>
    <au><snm>Mahmassani</snm><fnm>HS</fnm></au>
  </aug>
  <source>Transportation Research Part C: Emerging Technologies</source>
  <publisher>Elsevier</publisher>
  <pubdate>2015</pubdate>
  <volume>59</volume>
  <fpage>375</fpage>
  <lpage>-390</lpage>
</bibl>

<bibl id="B46">
  <title><p>Clustering of vehicle trajectories</p></title>
  <aug>
    <au><snm>Atev</snm><fnm>S</fnm></au>
    <au><snm>Miller</snm><fnm>G</fnm></au>
    <au><snm>Papanikolopoulos</snm><fnm>NP</fnm></au>
  </aug>
  <source>IEEE Transactions on Intelligent Transportation Systems</source>
  <publisher>IEEE</publisher>
  <pubdate>2010</pubdate>
  <volume>11</volume>
  <issue>3</issue>
  <fpage>647</fpage>
  <lpage>-657</lpage>
</bibl>

<bibl id="B47">
  <aug>
    <au><cnm>{FHWA}</cnm></au>
  </aug>
  <source>\url{https://ops.fhwa.dot.gov/publications/tt_reliability/TTR_Report.htm}</source>
  <pubdate>2019</pubdate>
  <note>[Online; accessed September-2019]</note>
</bibl>

</refgrp>
} 



\newpage

\section*{Figures}

  \begin{figure}[h!]
  \centering
  \includegraphics[width=0.9\linewidth]{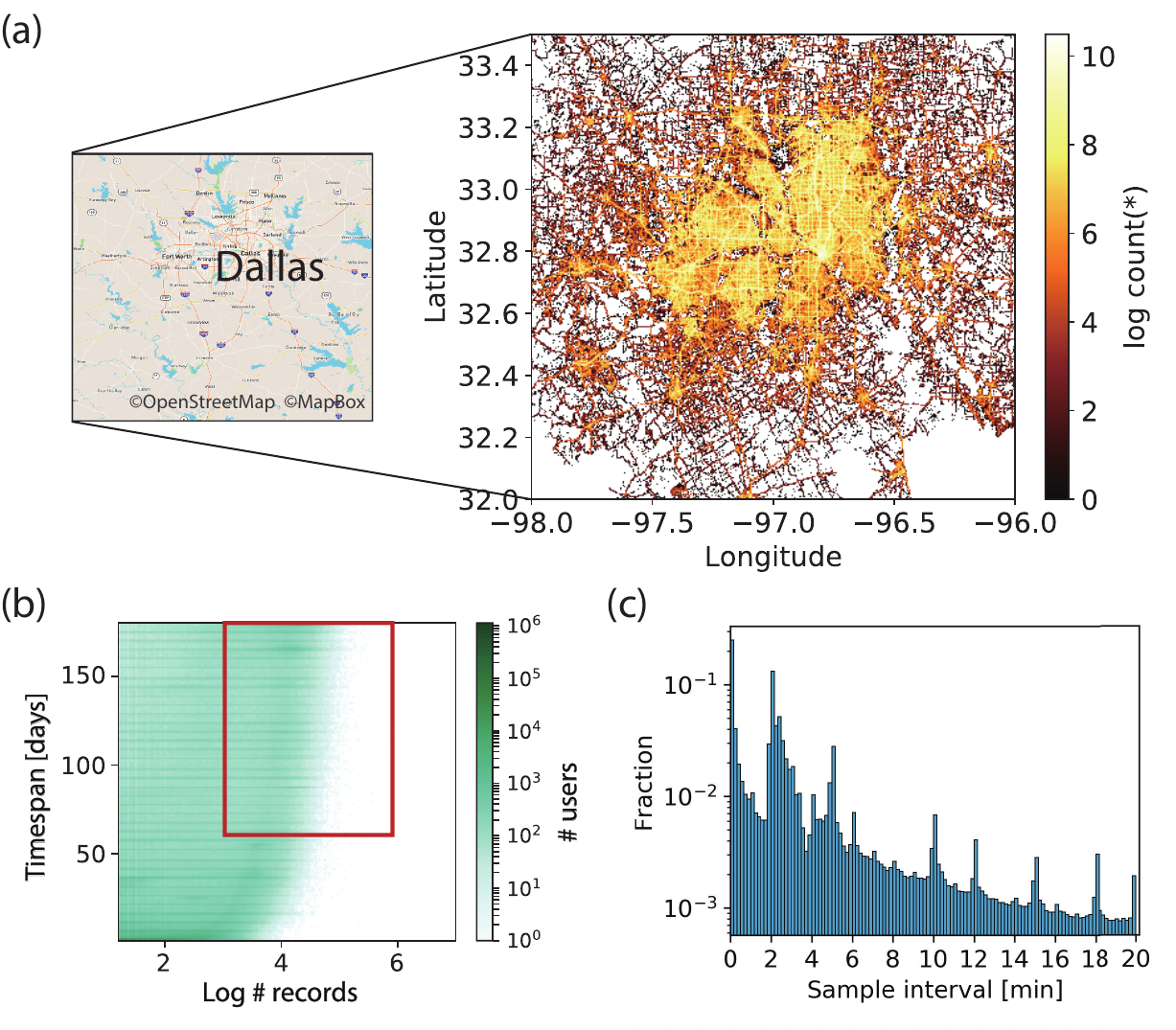}
  \caption{\csentence{LBS data and user selection.}
      (a) Spatial distribution of users' traces in the LBS data, measured by the logarithm of the total visitation in each grid. (b) User timespan versus his/her number of records. Users outside the red rectangle are eliminated in further analysis. (c) Distribution of sample interval of LBS data.}
      \label{fig:data}
      \end{figure}

\begin{figure}[h!]
\centering
\centerline{\includegraphics[width=0.90\linewidth]{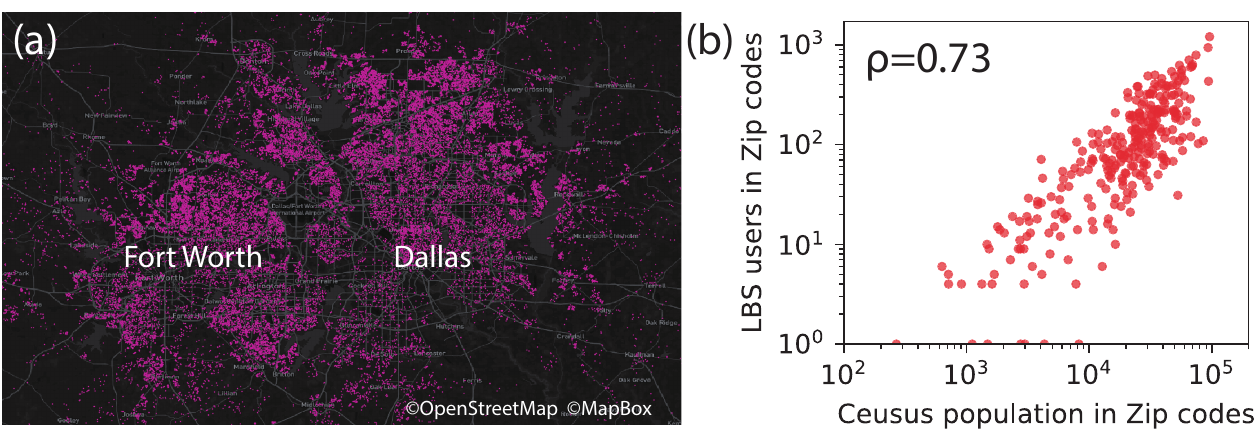}}
  \caption{\csentence{User home location estimation.}
      (a) User home locations estimated with visitation time and frequency. (b) Comparison between the active LBS users settling in the ZIP codes and the corresponding population from the census data.}
      \label{fig:home}
      \end{figure}
      
 \begin{figure}[h!]
\centering
\centerline{\includegraphics[width=0.90\linewidth]{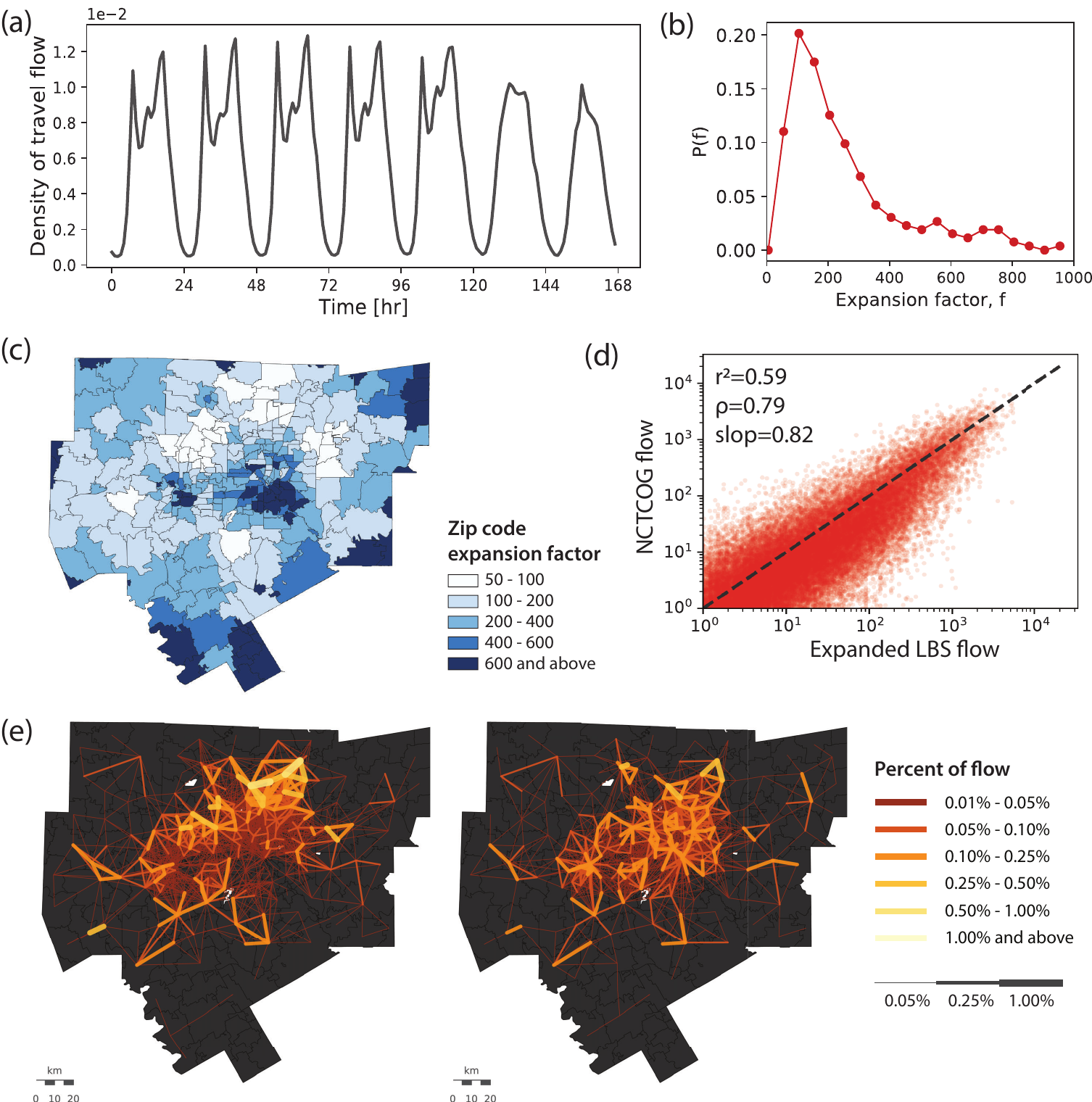}}
  \caption{\csentence{Validation of travel demand generated by LBS data.}
      (a) Fraction of travel flow per hour during one week generated by LBS data. (b) Distribution of expansion factor at Zip code level. (c) Visualization of expansion factor of each Zip code. (d) Comparison of vehicular travel flow during the morning peak hours between expanded LBS data and the NCTCOG data. Each red point represents an OD pair at ZIP code level. (e) Visualization of vehicular travel flow above 0.01\% between ZIP codes during the morning peak hours generated by expanded LBS data (left) and the NCTCOG data (right).}
      \label{fig:demand}
      \end{figure}
      
\begin{figure}[h!]
\centering
\centerline{\includegraphics[width=0.9\linewidth]{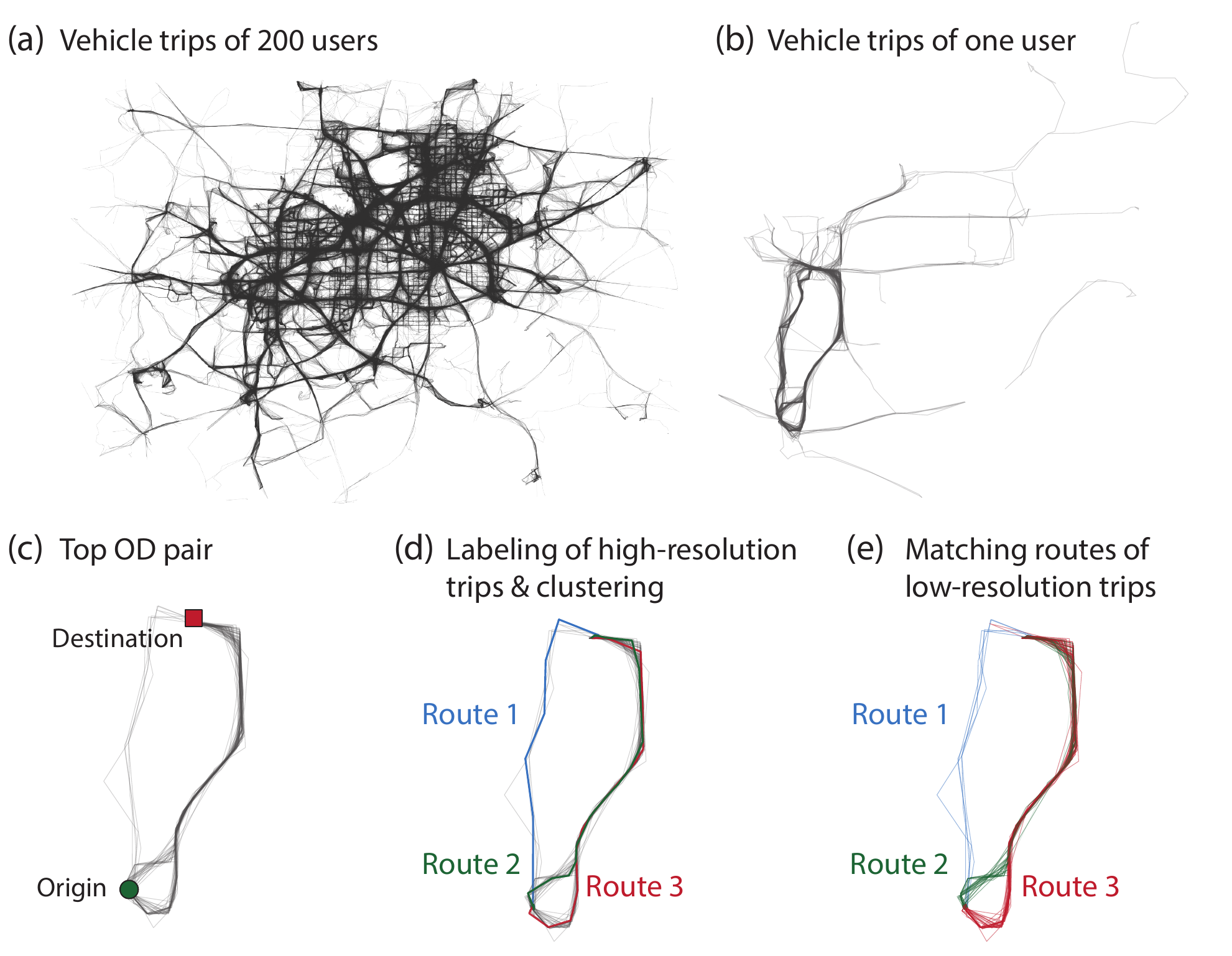}}
\caption{\csentence{Route detection from the routine OD pairs.} (a) All vehicle trips of 200 sample users. (b) All vehicle trips of one randomly selected user. (c) Trips in the top OD pair. (d) Routes detected from the high-resolution trips. (e) Matching the routes of low-resolution trips.}
\label{fig:routes}
\end{figure}

\begin{figure}[h!]
\centering
\centerline{\includegraphics[width=0.9\linewidth]{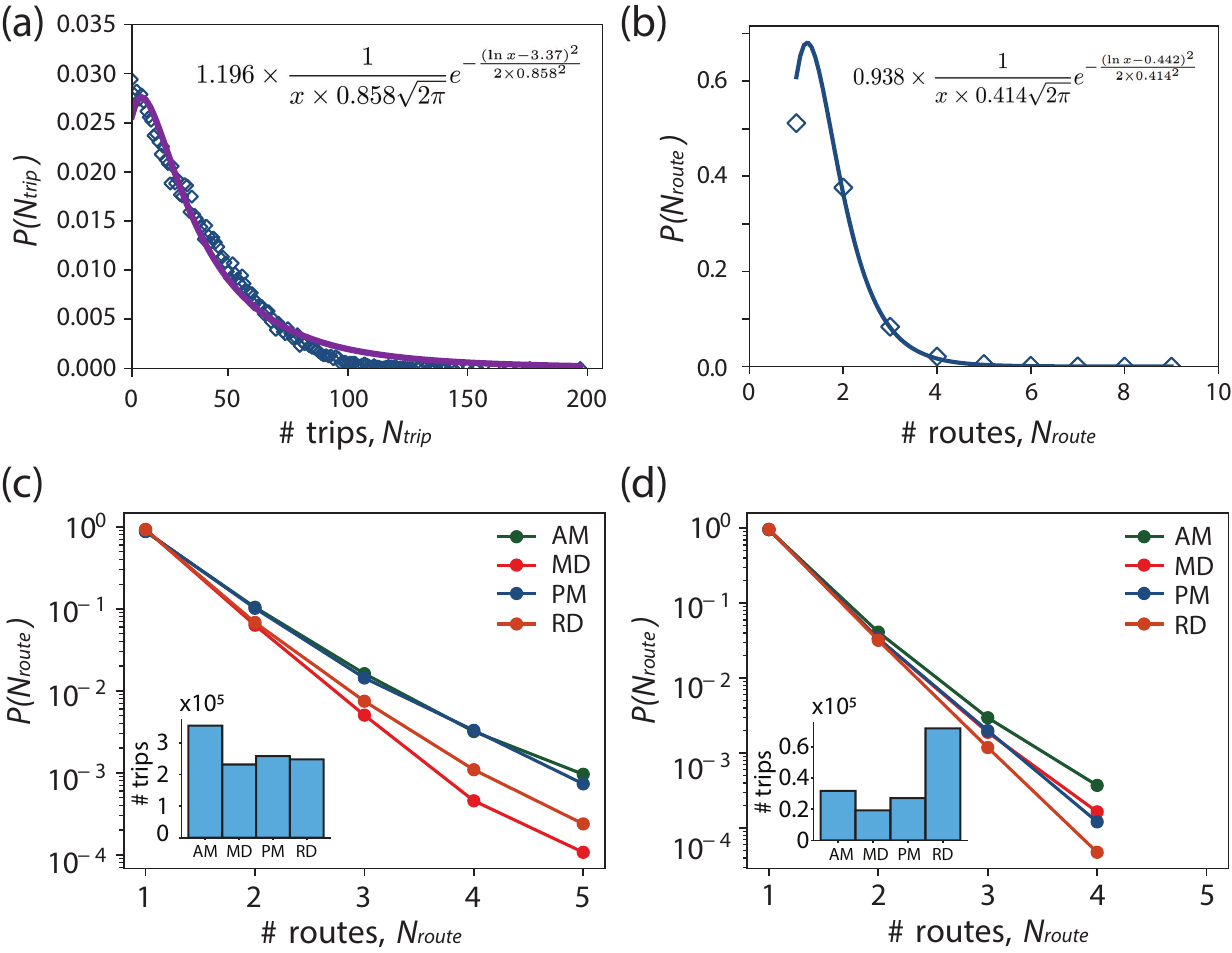}}
\caption{\csentence{Route choice behavior analysis.} (a) Distribution of number of trips, $N_{trip}$, in the routine OD pairs for all active users in the LBS data. The data follows a log-normal distribution. (b) Distribution of number of routes, $N_{route}$, in the routine OD pairs, also follows a log-normal. (c) Distribution of $N_{route}$ during different time periods, e.g., AM, MD, PM, and RD, on weekdays. The inset shows the number of trips during each time period on weekdays. (d) Distribution of $N_{route}$ during different time periods on weekends. The inset shows the number of trips during each time period on weekends. }
\label{fig:distribution}
\end{figure}

\begin{figure}[h!]
\centering
\centerline{\includegraphics[width=0.9\linewidth]{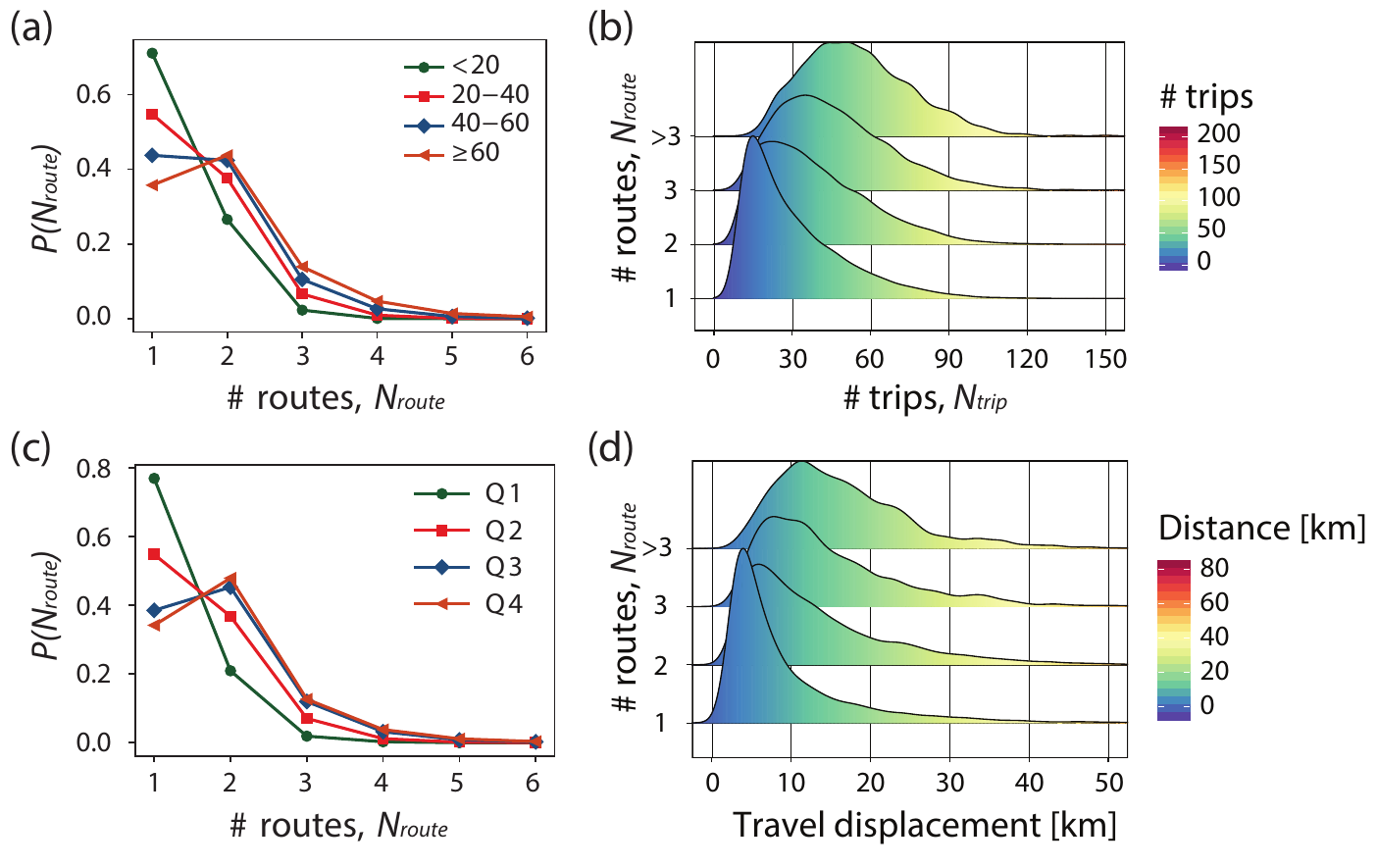}}
\caption{\csentence{Connection between number of routes and the number of trips and travel displacement.} (a) Fraction of $N_{route}$ for users grouped by number of trips. (b) Distribution of $N_{trip}$ of travelers with different $N_{route}$, e.g., one, two, three, and more than three routes. (c) Fraction of $N_{route}$ for users grouped by range of travel distance. (d) Distribution of user travel displacement for travelers with different $N_{route}$.}
\label{fig:groups}
\end{figure}

\begin{figure}[h!]
\centering
\centerline{\includegraphics[width=0.9\linewidth]{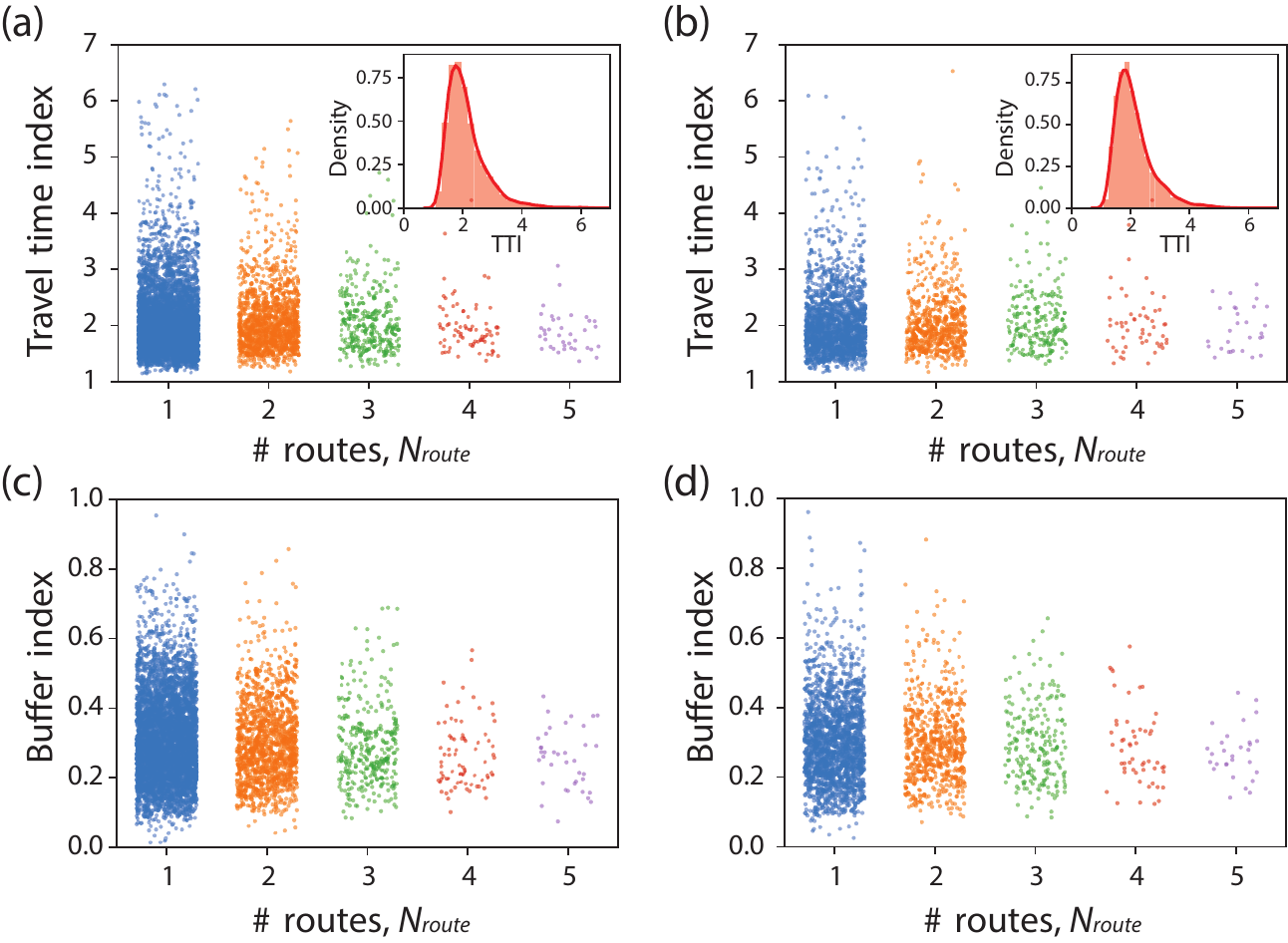}}
\caption{\csentence{Travel time index and buffer index of users taking different number of trips.} (a) Travel time index of the users with different number of routes during AM peak hours (7:00-10:00) on weekdays. The inset presents the distribution of TTI of all users during AM peak hours. (b) Travel time index of the users with different number of routes during PM peak hours (16:00-19:00) on weekdays. The inset presents the distribution of TTI. (c) Buffer index of the users with different number of routes during AM peak hours on weekdays. (d) Buffer index of the users with different number of routes during PM peak hours on weekdays.}
\label{fig:traveltime}
\end{figure}



\section*{Tables}

\begin{table}[h!]
\begin{center}
\caption{Mean values and STD of travel time index and buffer index of users taking different number of trips. The TTI$^*$ and BI$^*$ indicate the values mean values calculation are selected in the $95\%$ confidence interval.}
\label{tab:traveltime}
\begin{tabular}{l|ccccc}
\toprule
 \textbf{\# routes} & \textbf{1} & \textbf{2} & \textbf{3}& \textbf{4}& \textbf{5}\\
\midrule
average TTI (AM peak) & 2.110 & 2.139 & 2.095 & 1.991 & 1.878 \\
average TTI$^*$ (AM peak) & 2.042 & 2.072 & 2.050 & 1.940 & 1.847 \\
STD of TTI (AM peak) & 0.669 & 0.654 & 0.556 & 0.468 & 0.341 \\ 
average TTI (PM peak) & 2.077 & 2.115 & 2.104 & 2.059 & 1.952 \\
average TTI$^*$ (PM peak) & 2.009 & 2.044 & 2.060 & 1.998 & 1.936 \\
STD of TTI (PM peak) & 0.654 & 0.696 & 0.524 & 0.552 & 0.389 \\ 
\midrule
average BI (AM peak) & 31.46\% & 30.50\% & 28.97\% & 26.67\% & 25.49\% \\ 
average BI$^*$ (AM peak) & 30.89\% & 29.91\% & 28.23\% & 26.03\% & 25.50\% \\ 
STD of BI (AM peak) & 12.84\% & 12.04\% & 11.52\% & 9.57\% & 8.50\% \\ 
average BI (PM peak) & 30.63\% & 30.41\% & 29.83\% & 28.84\% & 27.12\% \\
average BI$^*$ (PM peak) & 29.99\% & 29.72\% & 29.33\% & 28.33\% & 26.82\% \\ 
STD of BI (PM peak) & 12.52\% & 11.87\% & 11.00\% & 10.76\% & 7.22\% \\ 
\bottomrule
\end{tabular}
\end{center}
\end{table}


\end{backmatter}
\end{document}